%% file: paper.tex
\documentclass[twocolumn]{aastex62}

\newcommand{\expval}[1]{\ensuremath{\langle #1 \rangle}}

\usepackage{soul, xcolor}
\usepackage{physics} 
\setstcolor{red}
\usepackage{multirow}
\usepackage{amsmath}
\usepackage[capitalise]{cleveref}
\usepackage{tabularx}

\newcommand{\simpleDS}{{\tt simpleDS}}

\newcommand{\rimez}{{\tt RIMEz}}
\defcitealias{HERA2020}{HC20}
\def\HOneClimit{\citetalias{HERA2020}}
\newcommand{\herapspec}{{\tt hera\_pspec}}

\usepackage[caption=false]{subfig}

\received{\today}
\revised{\today}
\accepted{\today}
\submitjournal{ApJ}

\shorttitle{HERA Phase I Validation}
\shortauthors{}

\begin{document}

\renewcommand{\equationautorefname}{Eq.}
\renewcommand{\sectionautorefname}{Sec.}
\renewcommand{\subsectionautorefname}{Sec.}
\renewcommand{\figureautorefname}{Fig.}

\title{Validation of the HERA Phase I Epoch of Reionization 21 cm Power Spectrum Software Pipeline}

\input{author-list}

\collaboration{HERA Collaboration}

\begin{abstract}
We describe the validation of the HERA Phase I software pipeline by a series of modular tests, building up to an end-to-end simulation.
The philosophy of this approach is to validate the software and algorithms used in the Phase I upper limit analysis on wholly synthetic data satisfying the assumptions of that analysis, not addressing whether the actual data meet these assumptions.
We discuss the organization of this validation approach,  the specific modular tests performed, and the construction of the end-to-end simulations.
We explicitly discuss the limitations in scope of the current simulation effort.
With mock visibility data generated from a known analytic power spectrum and a wide range of realistic instrumental effects and foregrounds,
we demonstrate that the current pipeline produces power spectrum estimates that are consistent with known analytic inputs to within thermal noise levels (at the $2\sigma$ level) for $k > 0.2 h{\rm Mpc}^{-1}$ for both bands and fields considered.
Our input spectrum is intentionally amplified to enable a strong `detection' at $k\sim 0.2 h{\rm Mpc}^{-1}$ -- at the level of $\sim 25\sigma$ -- with foregrounds dominating on larger scales, and thermal noise dominating at smaller scales.
Our pipeline is able to detect this amplified input signal after suppressing foregrounds with a dynamic range (foreground to noise ratio) of $\gtrsim 10^7$. 
Our validation test suite uncovered several sources of scale-independent signal loss throughout the pipeline, whose amplitude is well-characterized and accounted for in the final estimates.
We conclude with a discussion of the steps required for the next round of data analysis.

\end{abstract}


\section{Introduction} 
\label{sec:Introduction}

Measurement of the highly-redshifted 21 cm hyperfine transition of neutral hydrogen holds great promise as a probe of the Epoch of Reionization, as well as earlier and later epochs.
Because the power spectrum of 21 cm fluctuations must be measured 
in the presence of foregrounds which are $\sim10^5$ brighter (in temperature units) than the EoR signal, 
the level of precision required of every aspect of the analysis is extraordinarily high.
Because the line-of-sight power spectrum is measured using the spectral axis, it is critically important to avoid introducing additional spectral structure in the data during the analysis, as this can contaminate the 21 cm spectrum with foreground power. 
This is particularly a problem for interferometric measurements, which mix spatial and spectral structure \cite{datta09, Parsons2012}.
Inaccuracies may also be introduced by analysis choices which affect the amplitude of the desired signal relative to other portions of the data, e.g., biased estimators of the power spectrum or over-fitting of foreground models or calibration parameters.  
It is thus necessary to demonstrate the accuracy of the analysis both for individual steps in the analysis, and for the interconnected, complicated chain of analysis from raw data to power spectrum.  

A number of groups are currently seeking to detect the HI fluctuation signal from the EoR via the power spectrum.  
Current efforts include those of the Murchison Widefield Array \citep[MWA;][]{Tingay2013, Dillon2014, Ewall-Wice2016b, Beardsley2016, Barry2019b, Li2019, Trott2020}, 
the Low Frequency Array \citep[LOFAR;][]{vanHaarlem2013, Patil2017, Gehlot2018, Mertens2020}, the Long Wavelength Array \citep[LWA;][]{Eastwood2019}, and
the Hydrogen Epoch of Reionization Array \citep[HERA;][]{deboer2017hydrogen}.
Prior work also includes the 
the Giant Metre Wave Radio Telescope \citep[GMRT;][]{Paciga2013} and
the Donald C. Backer Precision Array for Probing the Epoch of Reionization \citep[PAPER;][]{Kolopanis2019}.

A persistent problem has been that the complexity of the measurement, combined with the novelty of analysis techniques, has created situations in which significant biases are created in the final power spectrum in ways which are not initially obvious.  
Due to these complications, 
there have been a number of limits that have 
required significant revision.
These include the GMRT (\citet{Paciga2011} as amended by \citet{Paciga2013}) and PAPER (\citet{Ali2015} as amended by \citet{Ali2018} and \citet{Cheng2018}).  
\citet{Liu2020} provides a good overview of many of these issues.

In response, an increased effort has been to made to explore the effects of choices in 21 cm analysis pipelines via simulation.  
These studies have attempted to isolate specific effects, for example,  sky-based calibration errors \citep{Barry2016, EwallWice2017, Mouri2019}, redundant calibration errors \citep{Orosz2019, Byrne2019}, instrumental coupling systematics \citep{Kern2019}, power spectrum estimation \citep{Cheng2018}, foreground modeling and subtraction \citep{Chapman2012_gmca, Mertens2018, KernAndLiu2020}, interferometric imaging \citep{Offringa2019_uvgridding}, the effect of RFI \citep{Wilensky2020}, and data inpainting \citep{Offringa2019_inpainting, Trott2020}.

Increasing effort has also gone into connecting these isolated studies into more complete end-to-end simulations of the pipelines. 
For example, the MWA team has two parallel pipelines \citep{Jacobs2016, Trott2016, Barry2019a}.  The reliability of the pipeline in recovering a mock power spectrum (but without including the effects of calibration) was tested in \citet{Beardsley2016} and more explicitly in \citet{Barry2019b} (Figure 8).  
The LOFAR limits published in \citet{Mertens2020} have had the method simulated in \citet{Mertens2018} and the effect of calibration considered in \citet{Mouri2019} and Mevius et al. in prep.

This paper details the current status of an end-to-end simulation effort for the HERA pipeline, as a companion paper to \citet{HERA2020} (hereafter \citetalias{HERA2020}) and specifically addresses the instrument configuration and systematic effects of HERA Phase I.   
Importantly, as will be expanded upon later, these validation tests are aimed at verifying the accuracy of the HERA Phase I pipeline under the intrinsic assumptions of the pipeline itself. 
Furthermore, they provide a reproducible framework with which to evaluate future HERA analysis pipelines and data releases.
These tests are in principle sufficient to avoid the algorithmic errors leading to revisions such as those in \citet{Cheng2018}.

The outline of the paper is as follows:  
Section \ref{sec:HERA} briefly describes the HERA instrument and the software pipeline we are attempting to validate.  
Section \ref{sec:Methods} explains the underlying philosophy of software development and organization of the validation effort, while section \ref{sec:SimulationComponents} outlines the simulation methods used for each individual portion of the pipeline, and results of isolated tests of those portions. 
Section \ref{sec:EndToEnd} then shows the results for the end-to-end pipeline simulation and a comparison with an independent method of estimating the power spectrum. 
We conclude with a discussion of lessons learned and next steps in Section \ref{sec:Conclusions}.

\section{The HERA Instrument and Software Pipeline}
\label{sec:HERA}

\subsection{The HERA Instrument}
\label{subsec:HERAInstrument}

HERA \citep{deboer2017hydrogen} is a dedicated instrument to measure the power spectrum of spatial fluctuations in the strength of the hyperfine signal of neutral hydrogen during the Epoch of Reionization and Cosmic Dawn.  The final instrument, currently under construction at the SKA South Africa site, will comprise a core of 320 14-meter parabolic dishes in a fractured hexagonal-close-pack configuration \citep{DillonAndParsons2016} with 30 outrigger antennas.  It will operate over the frequency range 50 - 250 MHz ($27 < z < 5$). 

\begin{figure}
    \centering
    \includegraphics[width=\linewidth]{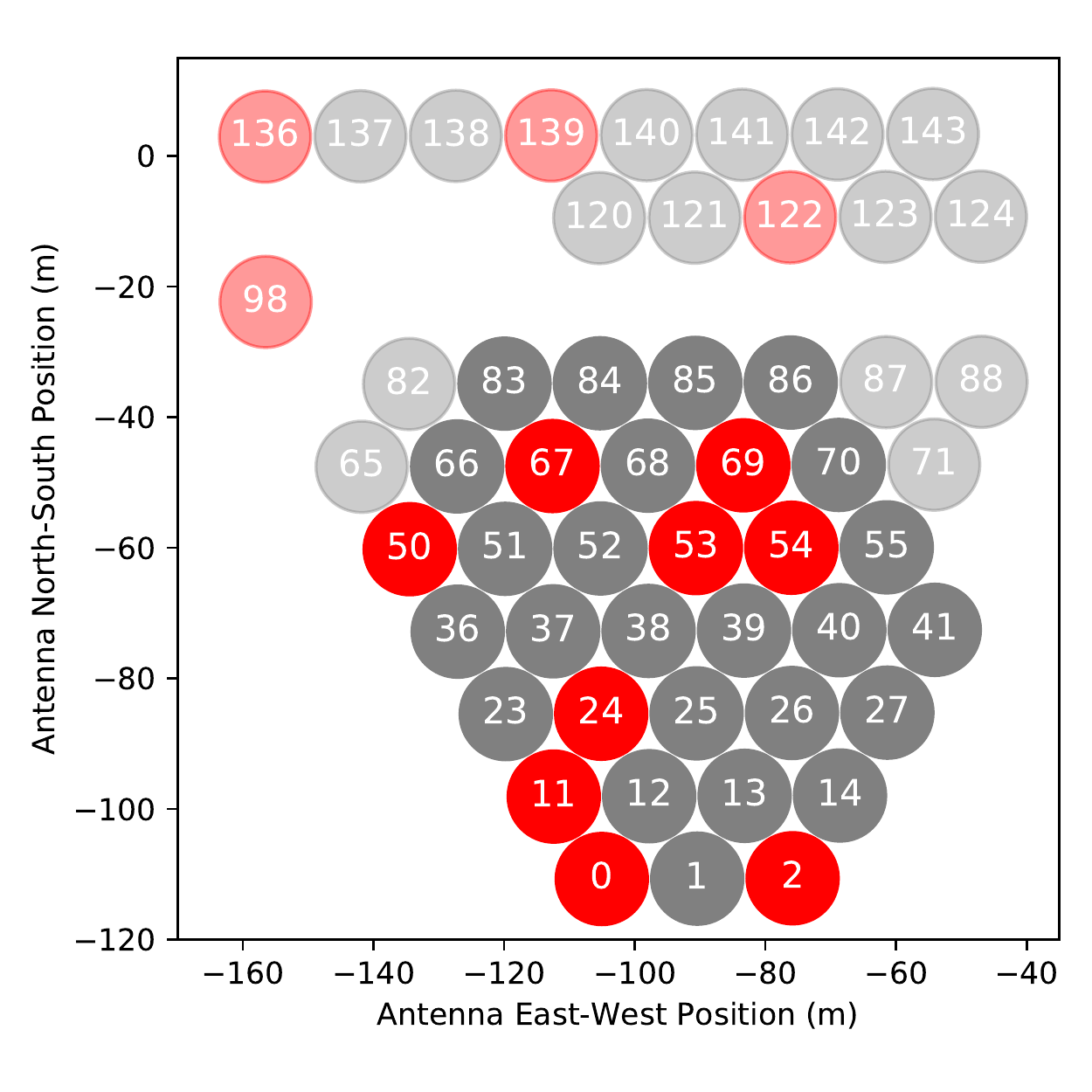}
    \caption{The array layout and antenna numbering scheme for the HERA Phase I data (c.f. Figure 2 of \HOneClimit).  The subset of antennas used in end-to-end validation are shown dark; the additional antenna present in the real array are transparent in this figure. Flagged antennas, shown in red, match those flagged in the real data \citep{Dillon2020}.}
    \label{fig:array_layout}
\end{figure}

Here we are concerned with the state of the instrument consistent with the \citetalias{HERA2020} data,\footnote{This dataset is referred to within the collaboration as H1C Internal Data Release (IDR) 2.2.} which comprises 39 active antennas operating from 100 - 200 MHz, using the feed type described in \citet{fagnoni19} in the configuration shown in Figure \ref{fig:array_layout}.  In particular, the systematic effects considered are specific to that instrument.  The Phase II instrument under construction has an entirely different feed \citep{fagnoni20}, signal path, and correlator, and thus will have very different behavior; the simulation and validation of the analysis pipeline for that instrument are the subject of future work. 

\subsection{Brief Overview of the HERA Analysis Pipeline}
\label{subsec:HERAPipeline}

\begin{figure*}
    \centering  
    \includegraphics[width=0.9\linewidth]{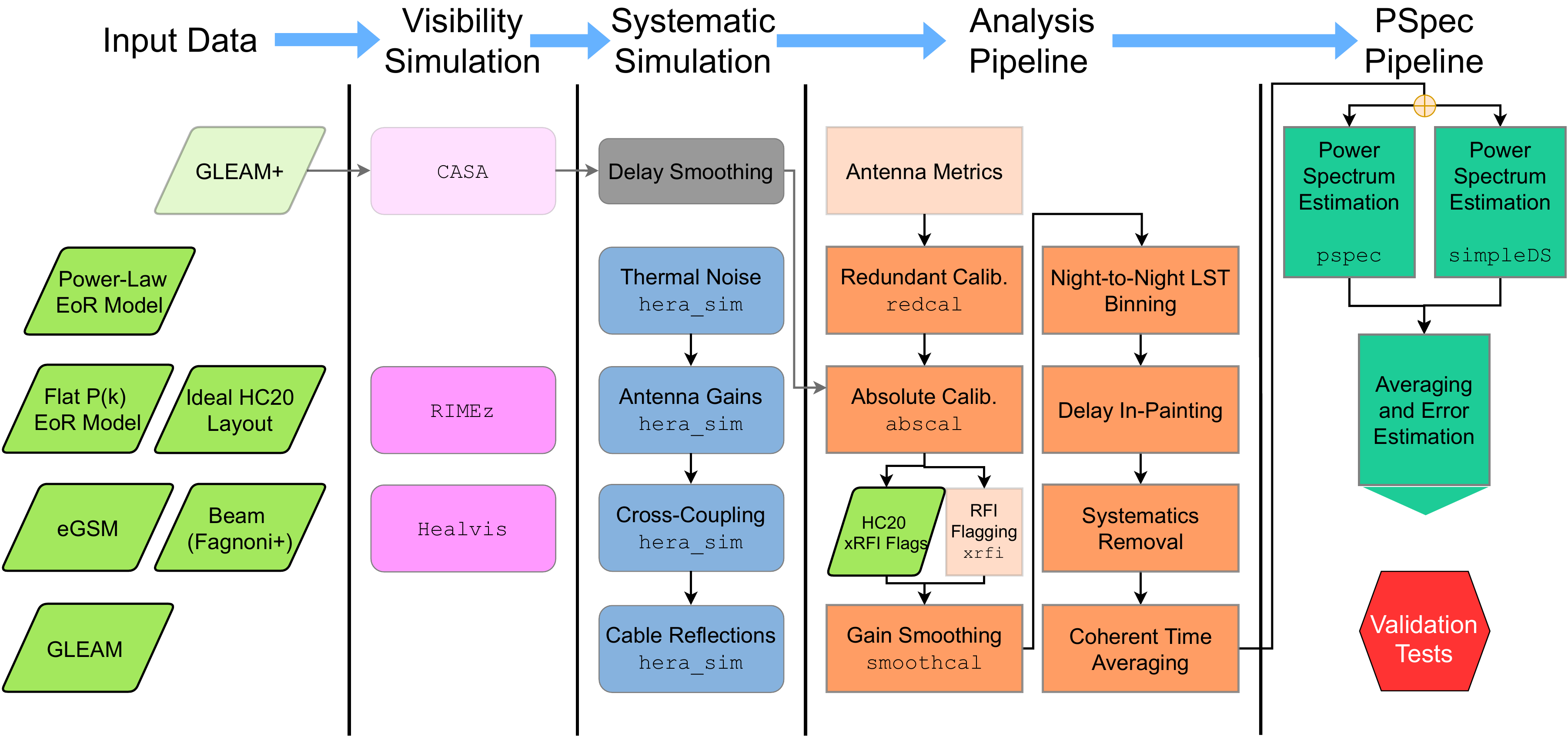}
    \caption{A schematic representation of the entire validation pipeline. See \cref{fig:validation_matrix} for the way in which individual components were tested. Simulation and analysis flows from left to right, and top to bottom where applicable, as indicated by the black arrows. Transparent nodes are not validated in the validation pipeline for H1C IDR2.2, but are included here for clarity. Overlay of the input H1C IDR2.2 flags in place of \textsc{xrfi} indicates that we did not simulate and extract RFI in the validation pipeline, but rather injected pre-existing real flags from the data. Absolute calibration (\textsc{abscal}) requires an input high-fidelity model of the sky. In the real analysis pipeline, this model was produced with CASA from observed sky visibilities. Grey arrows indicate this pipeline.  However, in our pipeline, we use the exact simulated foreground model -- with delay smoothing applied. }
    \label{fig:validation_pipe}
\end{figure*}

The HERA data reduction pipeline takes raw data output from the correlator and delivers calibrated visibilities with bad antennas, RFI, and other anomalies flagged or removed, and produces delay-type \citep{Parsons2012_dspec, Morales2019} power spectra, with accompanying error bars and null tests. A full accounting is given in \citetalias{HERA2020}, and the key steps are shown in Figure \ref{fig:validation_pipe} (in the ``Analysis Pipeline'' and ``PSpec Pipeline'' columns).  Steps in the pipeline which were not included in the validation are indicated with faded colors.  We briefly outline the steps here, indicating the reason certain steps were not included, and where further information can be found.

Raw data is delivered from the correlator (not shown in Fig. \ref{fig:validation_pipe}, but equivalent to the output of the systematic simulation).  
A first round of quality checking attempts to identify antennas which are not performing correctly (``Antenna Metrics''), as inclusion of these in the following step (``Redundant Calibration'') adversely affects the results.  
We did not attempt to produce simulations of the various kinds of antenna defect (hardware failure, incorrect wiring), but simply excluded a subset of possible antennas to simulate the effects of flagging the antennas.  
``Redundant calibration'' uses the constraint of the repeated array configuration to solve for internal degrees of freedom by forcing identical baselines to have identical response.  
The particular implementation is described in greater detail in \citet{Dillon2020}.  
``Absolute Calibration'' solves for the remaining degenerate parameters undetermined by redundant calibration.
This is done by comparing the redundantly calibrated data to a set of model visibilities which have been absolutely calibrated.
Those model visibilities had their absolute flux and phase determined by CASA, using a model of GLEAM sources, as described in \citet{Kern2020_abscal}.
The model visibilities are then subsequently smoothed to not contain structure beyond the baseline's horizon delay (with an additional 50\,ns buffer) or 150\,ns, whichever is larger.
Using a calibration based on an incomplete sky model is known to produce biases in the estimated power spectra \citep{Barry2016}.
This is mitigated for HERA because the sky-based calibration is only used to determine the degenerate parameters (fewer degrees of freedom) and also because of the subsequent gain smoothing.
Because we do not know the level to which the CASA model was actually incomplete, we do not simulate the effect of that error in this analysis, in keeping with our philosophy of simulating data which respects the pipeline assumptions.  
We discuss this as part of future work in \autoref{sec:Conclusions}.

At this point the data are flagged (``RFI Flagging'') based on a number of metrics, including the output of the calibration steps, with the goal of removing particularly non-redundant behavior and RFI.  
An entirely separate study is required to understand the efficacy of this algorithm in removing RFI, which we defer to later work.  
While unflagged RFI is a concern for the power spectrum \citep[e.g.,][]{Wilensky2020}, we show in \HOneClimit\ that there is no strong evidence for unflagged RFI at the current noise levels, and so here we concern ourselves primarily with the effect of gaps in the data resulting from RFI flagging.  Thus there is not a simulated RFI injection and removal step; we simply copy the flagging pattern from the real data and use it to create gaps in the simulated data.  

Following this, the final gains are smoothed in frequency and time to avoid imparting spurious structure (``Gain Smoothing''), and a final calibrated dataset is produced for each night.  The data are then averaged over nights of observation (``LST Binning''), with data taken at the same LST for a given baseline averaged together.  
Any remaining gaps in the averaged baseline visibilities  due to flagged data are filled (``Delay Inpainting'')  \cite[cf. \S\ref{sec:Methods:Flagging} ;][]{Parsons2009,Kern2020_systematics}.  
At this point two systematic effects are corrected (``Systematics Removal''): the presence of internal reflections, causing an ``echo'' of the signal at different time delays, effectively changing the gain solutions; and a cross-coupling between antennas creating an additive signal.  
These effects are described in \citet{Kern2019,Kern2020_systematics}.  
Baselines are then averaged in time (``Coherent Time Averaging'') to produce the dataset used in ``Power Spectrum Estimation'' (the pipeline used in \HOneClimit\ uses \herapspec, which we validate against an alternative power spectrum pipeline -- \texttt{simpleDS} -- in this study).  
Additional averaging occurs over different baseline types and the ``cylindrical'' average from $(k_\parallel,k_\perp)$ to $k$ to produce 1-D power spectra with associated errors.  
Further details about the pipeline can be found in \HOneClimit.  

\begin{table*}
    \centering
    \begin{tabularx}{\linewidth}{ll>{\raggedright\arraybackslash}p{8.5cm}}
        \hline
        \hline
        \textsc{Name} & \textsc{URL} [\url{https://github.com/}] & \textsc{Description} \\
        \hline
        \hline
        \textsc{Pipeline} & & \\
        \hline
        \texttt{hera\_cal} & \url{hera-team/hera_cal} & Redundant and sky-based calibration routines. \\
        \herapspec & \url{hera-team/hera_pspec} & Robust foreground-avoidance power spectrum \ (and covariance) estimation.\\
        \hline
        \textsc{Simulation} & & \\
        \hline
        \rimez & \url{upenneor/rimez} & Fast and accurate visibility calulation implementing multiple methods for different source and beam function definitions.\\
        \texttt{spin1\_beam\_model} & \url{upenneor/spin1_beam_model} & Harmonic space decomposition of the HERA primary beam. \\
        \texttt{healvis} & \url{rasg-affiliates/healvis} & Fast visibility simulation based on \textsc{HEALPix} discretization.\\
        \texttt{pyuvsim} & \url{RadioAstronomySoftwareGroup/pyuvsim} & Accurate visibility simulation of point sources with very limited approximations. \\
        \texttt{gcfg} & \url{zacharymartinot/redshifted_gaussian_fields} & Consistently simulate cosmological Gaussian fields over the full sky\\
        \texttt{hera\_sim} & \url{hera-team/hera_sim} & Add HERA-specific instrumental systematics to visibilities. \\
        \hline
    \end{tabularx}
    \caption{Table of repositories tested in this validation effort.}
    \label{tab:repos}
\end{table*}

\section{Methods} 
\label{sec:Methods}

\subsection{Overview of Validation Effort} \label{sec:meth:overview}

The desire to ensure that the HERA analysis pipeline does not produce biased results motivated the creation of a separate ``Validation'' group within HERA (see Appendix \ref{sec:ValidationSubsystem}), which seeks to provide an ongoing framework for testing the pipeline via realistic simulations.  The scope of this paper is somewhat more narrow.
The HERA analysis and power spectrum pipeline described above is clearly a large and complex system. 
It is implemented, in part, by the public software repositories in \cref{tab:repos}, which 
comprise at least four complete, original, python packages, with upwards of 40,000 standard lines of code between them.
While each of the packages is written to a high collaboration standard (see Appendix \ref{sec:CodeStandards} for details), the interplay between the sub-components is much more difficult to test.  
What we wish to do here is verify that the software pipeline used in \HOneClimit\ performs as expected {\it in the case that the simulated data match the underlying assumptions of the analysis.} 
Importantly, we do {\it not} explore the effects of violating certain key assumptions of the pipeline, including perfect redundancy of antenna elements \citep{Dillon2020}, or systematic effects which differ substantially from \citet{Kern2019, Kern2020_systematics}. 
That is, we assume that the physical effects for which the analysis pipeline was designed to remove are the only (non-negligible) effects in the data, and that the modelling of these effects in the pipeline is comprehensive in principle.
(In some cases it was not possible to completely simulate what was done in \HOneClimit\, and we have noted this.)
Our key metric for this validation is the recovery of a known power spectrum, without significant bias in the recovered signal, at the level of error bars that are consistent with the known level of thermal noise and its coupling to the signal, following the error analysis in \citet{tan2020}.

The approach used here tests sub-components of the analysis with multiple simulations but does not attempt a statistical suite of simulations of the full end-to-end result.  This is partially the result of practical limitations of computation (many aspects of the simulation pipeline would need to be sped up), but also because we expect (and show) that in the absence of systematic effects which do not deviate from our assumptions, the output power spectrum is reproduced within the errors.  A more thorough investigation of ensemble effects is appropriate as the limits continue to come down, and in the exploration of the violation of pipeline assumptions.

At the current sensitivity of HERA, we do not expect to make high-significance detections of the EoR power spectrum. 
Consequently, our criterion for ``how good is good enough?'' in assessing the results of our simulations is that any systematic effect in the analysis are smaller than the expected error bars on the EoR spectrum, or the systematic errors in its calibration.  
In practice, that means we consider effects ``small'' if they cause a change in the power spectrum of less than $10\%$.
This bound will clearly need to be tightened as we begin to move toward detections.
It is worth commenting that errors may appear at different points in the analysis, and may affect the calibration gains, the individual visibilities, or the power spectrum itself.  
Our metric is the power spectrum, and we note that errors in gains $g$ typically propagate to the power spectrum as $g^4$, and errors in visibilities $V$ as $V^2$.  
Thus errors which affect the gains or visibilities must meet correspondingly smaller fractional error requirements so that the final effect on the power spectrum stays within the desired bound.

\subsection{Schematic Overview of the Validation Effort} 
\label{sec:meth:steps}

\begin{figure*}
    \centering
    \includegraphics[width=0.9\linewidth]{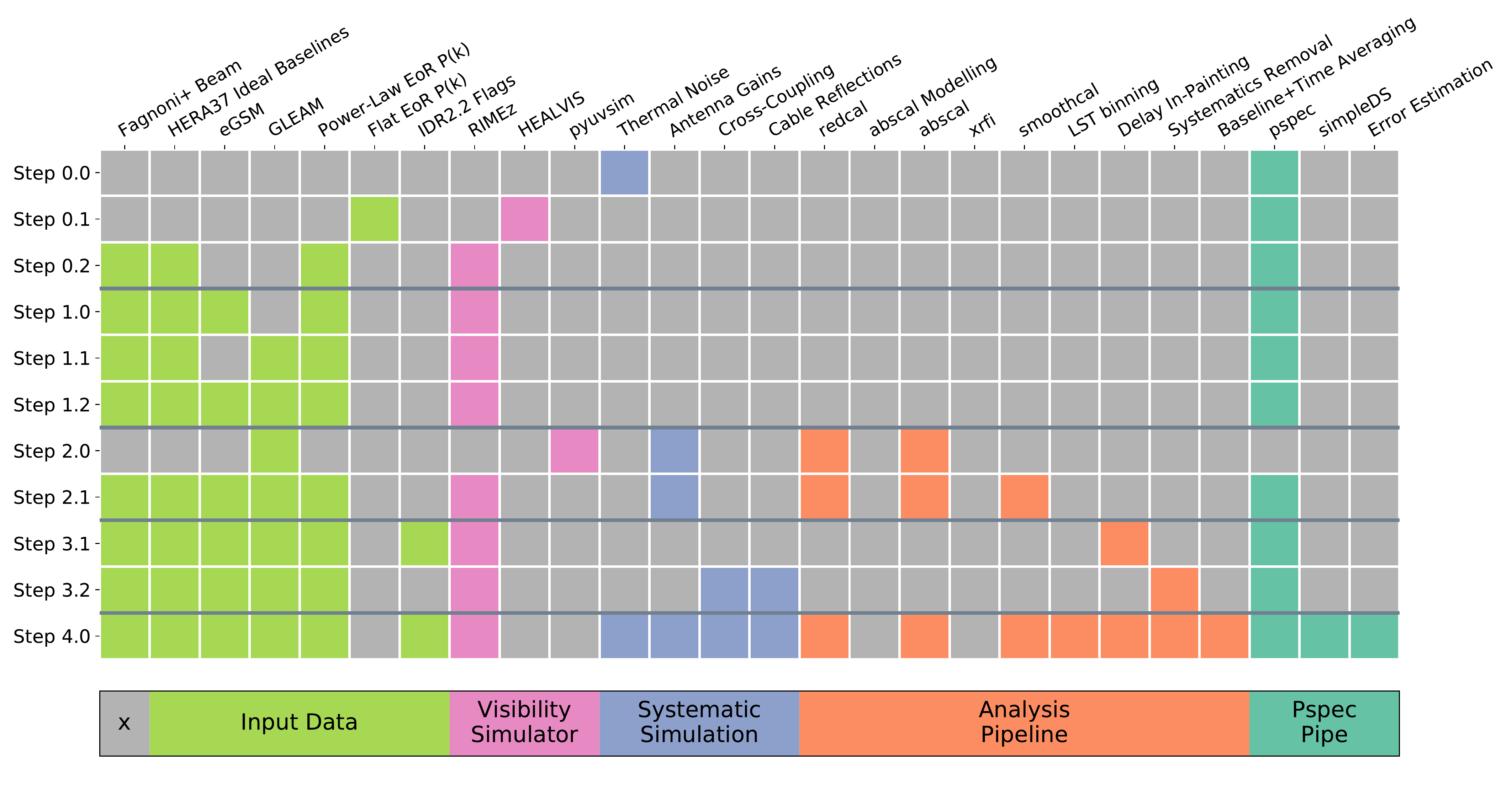}
    \caption{Components included in each validation step. Thick horizontal lines represent boundaries between major steps. Varying colors represent different subcategories of the pipeline. Where applicable, components are simulated/applied from left to right. See \cref{fig:validation_pipe} for a more detailed flow diagram of the simulation and analysis components and products.}
    \label{fig:validation_matrix}
\end{figure*}

We designed the validation effort to be incremental, building complexity in successive steps, and finally resulting in a simulation including a large fraction of the physical effects that the HERA pipeline attempts to address.
We divided the various components required for a full simulation into steps and tested each.  
The various simulation components are outlined in 
\autoref{fig:validation_pipe}, and the testing steps 
in 
\autoref{fig:validation_matrix}.
Section \ref{sec:SimulationComponents} describes these in more detail.

A row of \cref{fig:validation_matrix} indicates which elements of the simulation were included in the step.
For each of these steps (except \verb|2.0|) the primary metric of success involves the estimated power spectrum. 
As we progress through the steps, generally more elements are included, i.e.,
these are \textit{integration} tests where we cumulatively test interaction between components.
This has the potential to uncover undetected errors concerning the interaction of individual components, but also the potential to hide errors that are negligible in the final power spectrum metric.

Each specific \verb|major.minor| step tests a unique pathway through \cref{fig:validation_pipe}, combining different inputs, simulator and systematics with relevant pipeline steps. \cref{fig:validation_matrix} encodes each of these pathways in matrix form. Note that the steps were \textit{not} meant to test every possible combination of components, but rather to coherently build towards the ultimate test, which essentially combines all of them.
The reason for this incremental complexity building was pragmatic -- defining a full end-to-end simulation containing all known physical effects is a large task, and it is not necessarily clear from the outset what form an effect will have on the final result.
Furthermore, in the case that the test fails, it is difficult to disentangle effects and determine which component (or combination thereof) caused the failure. 
Increasing complexity gradually builds confidence in the individual analysis components before combining them.

\section{Simulation Components}
\label{sec:SimulationComponents}

We now walk sequentially through the various components of the simulation, as shown in the first three columns of \autoref{fig:validation_pipe},
and describe how they are constructed, including the methods and codes (\autoref{tab:repos}), as well as how they were tested within the rubric of \autoref{fig:validation_matrix}.
With any simulation, there are physical and instrument effects which are neglected, either due to ignorance (i.e., unknown systematic effects) or practical limitations in including them in the simulation.  We have included in \cref{fig:validation_pipe} a set of effects which encapsulate the most complete description of the sky and instrument which we are able to construct at this time.  There are known additional effects which are the subject of future work, which we discuss further in \autoref{sec:Conclusions}.

\subsection{Instrument and Foreground Models}

The first step in simulating instrumental output is to make models of the desired signal (``mock EoR'') and astrophysical foregrounds (point sources and diffuse emission), as well as to simulate the antenna response pattern and interferometer layout.

\subsubsection{Mock EoR}
\label{sec:mock_eor_step0}

To be able to test the unbiased recovery of an EoR power spectrum, it was highly desirable to produce a mock EoR sky with a known, analytic power spectrum $P(k)$.
It was deemed more important that the power spectrum be analytically known and that the simulated EoR be full-sky, covering both $4\pi$ steradians and the full observation bandwidth, than that it be derived from a physical simulation. 
While this means that the mock EoR we inject will not have the most realistic power spectrum, nor will it have a non-Gaussian component, these are second-order effects for ascertaining whether there is bias in the recovery of the power spectrum.

We chose to use for our mock-EoR signal a realization of a Gaussian random brightness temperature field $T_e(\va{r}, z)$ (expressed in the emitted frame) with a power spectrum with a {\it shape}
\begin{align}
\label{eq:AnalyticPk}
P_e(k, z) = A_0 k^{-2},
\end{align}
that approximates power spectra obtained from cosmological simulations.  
The observed field is given by
\begin{align}
    T( \vu{s}, \nu) = \frac{\nu}{\nu_e} T_e(\mathbf{r} = r_\nu \vu{s}, z = z_\nu)
\end{align}
where $\nu$ is the observed frequency, $\nu_e$ is the rest frequency, $z_\nu = \frac{\nu_e}{\nu} - 1$ is the redshift of the source point in the  direction $\vu{s}$ on the sky and 
\begin{align}
r_\nu = c \int_0^{z_\nu} \frac{\dd{z}}{H(z)}
\end{align}
is the comoving distance in terms of the Hubble function $H(z)$.  Note that the cosmological parameters are implicit in the comoving distance $r_\nu$; we have used the same parameters as in subsequent power spectrum estimation.

We can create realizations of this temperature field by expanding in spherical harmonic modes
\begin{align}
T(\vu{s}, \nu) = \sum_{\ell=0}^\infty \sum_{m=-\ell}^m a_{\ell m}(\nu) Y^*_{\ell m}(\vu{s})
\end{align}
and generating $a_{\ell m}$ that satisfy
\begin{align}
\expval{a_{\ell m}(\nu) a^*_{\ell' m'}(\nu') } = C_\ell(\nu, \nu') \delta_{\ell \ell'} \delta_{m m'}.
\end{align}
The cross-frequency angular power spectrum $C_\ell(\nu, \nu')$ is related to the original power spectrum $P_e(k)$ by
\begin{align}
C_\ell(\nu, \nu') & = \frac{2}{\pi} \frac{\nu \nu'}{\nu_e^2} \int_0^\infty \dd{k} k^2 j_\ell(r_\nu k) j_\ell(r_\nu' k) P_e(k).
\end{align}

For our chosen form of the power spectrum, \autoref{eq:AnalyticPk}, this takes the simple form
\begin{align}
C_\ell(\nu, \nu') & = \frac{\nu \nu'}{\nu_e^2} \begin{cases}
\frac{A_0}{2 \ell + 1} \frac{r_\nu'^\ell}{r_\nu^{\ell+1}} & \text{ if } r_\nu' \leq r_\nu, \\
\frac{A_0}{2 \ell + 1} \frac{r_\nu^\ell}{r_\nu'^{\ell+1}} & \text{ if } r_\nu' > r_\nu. \\
\end{cases}
\end{align}
Our mock-EoR signal is then a realization of this cross-frequency spectrum.
The maximum $\ell$ necessary is determined by the effective angular band limit imposed by the natural spatial filtering of the simulated interferometric array and sufficient error control on the visibility calculation.

The final field on the sky is expressed as a specific intensity (in units of Jy / str) using the conversion
\begin{align}
\label{eq:KelvinToJansky}
I_\nu(\vu{s}, \nu) = 
\kappa(\nu) T(\vu{s}, \nu)
\end{align}
The conversion factor is given by
\begin{eqnarray}
\label{eq:KappaDef}
    \kappa(\nu) & = & {2 k_B} \frac{\nu^2}{c^2} \times 10^{26} \; \;
    \left[\mathrm{\frac{Jy/str}{K}} \right] \\
    & = & \frac{2 k_B}{A(\nu) \Omega(\nu)} \times 10^{26}
\end{eqnarray}
where $k_B$ is Boltzmann's constant in SI units and $A(\nu)$ and $\Omega(\nu)$ are the effective area and solid angle of of the beam, respectively.

\begin{figure}
    \centering
    \includegraphics[width=\linewidth]{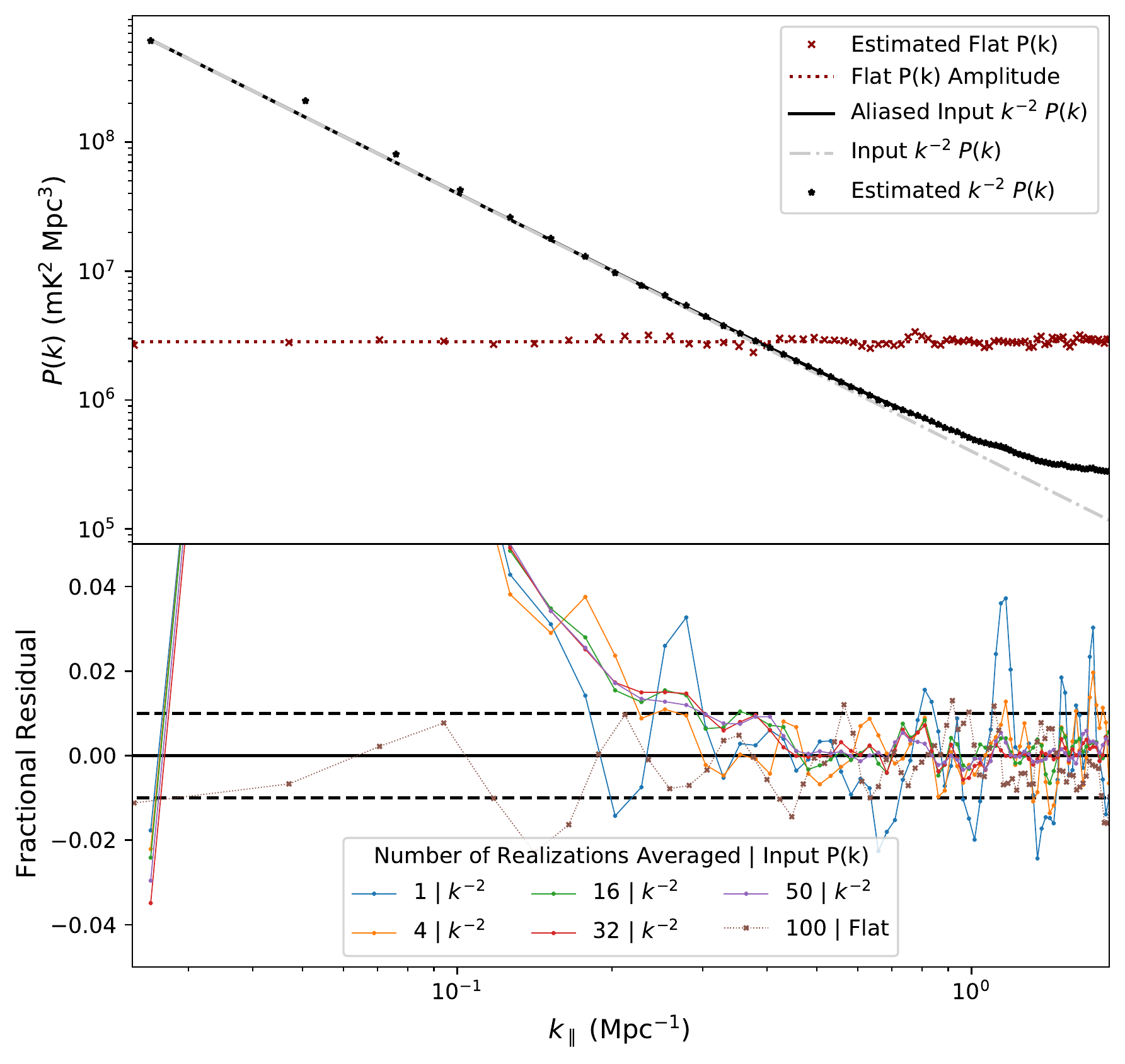}
    \caption{A baseline test of recovering the power spectrum for different power spectrum shapes.  Top: An analytic $P(k)\propto k^{-2}$ (orange) was converted to its corresponding $C_\ell(\nu,\nu')$; harmonic realizations of this correlation were input to \rimez\ to generate mock visibilities,  and the delay spectrum was estimated using \herapspec.  The results of a single realization are shown (blue), along with the calculated deviation due to the approximate window function of \autoref{eq:Aliasing}.  (The amplitude here is arbitrary, and not related to the level used in the end-to-end test (Section \ref{sec:EndToEnd}).)
    Bottom: the fractional deviation between the input power spectrum and the recovered one, after correction by \autoref{eq:Aliasing}, as a function of the number of realizations averaged.  Random fluctuations about the mean analytic form are expected due to cosmic variance; those fluctuations average down as shown. For $k \ge 0.2$, systematic deviations are $<1\%$.  The systematic deviations at low $k$ are due to not properly calculating the window function \autoref{eq:GeneralWindow}.  
    }
    \label{fig:PowerSpectrumValidation}
\end{figure}

To verify \herapspec's normalization conventions and cosmological conversions in going from visibilities to power spectra,   
we tested the recovery of $P(k) \propto k^{-2}$ in the absence of any foreground emission, noise, or instrumental corruption beyond the beam (designated Step 0.2 in \autoref{fig:validation_matrix}).
The results are shown in \autoref{fig:PowerSpectrumValidation}.
While the agreement is generally good, the results highlight an important aspect of the power spectrum measurement.
The estimated power spectrum $\hat{P}(k)$ is related to the true spectrum via a window function $W(k, k')$ via
\begin{align}
    \label{eq:WindowFunction}
    \expval{\hat{P}(k)} & = \int_{0}^{\infty} W(k, k') P(k') \dd{k'}
\end{align}
In general, $W(k, k')$ is complicated, and cannot be made equal to the ideal $\delta(k - k')$.
A discussion of the window function is included in \autoref{sec:WindowFunction}, and 
a general expression is given in \autoref{eq:GeneralWindow}.  
A full calculation of the window function would naturally include effects such as the curvature of the sky \citep[e.g.,][]{Liu2016spherical} and the bandwidth and resolution of the frequency sampling of the data.  
The window function computed by \herapspec\ (and given in \HOneClimit, Equation 19) does not fully implement \autoref{eq:WindowFunction} and consequently suffers from small biases at low and high $k$.
In \cref{fig:PowerSpectrumValidation}, we show the size of these biases.  In the range where we are most sensitive ($0.2 \le k \le 0.5 ~h~\mathrm{Mpc}^{-1}$) these biases are intrinsically less than 1\%.  With a simple approximation the aliasing effect (\autoref{eq:AliasedPk}), the bias for $k > 0.5$ can be reduced to a similar level. 
The bias for $0.03 < k < 0.2$ due to the window function is more severe.  Part of the discrepancy is simply due to nearly uniform width windows in $k$ when integrated against a power law, but in general the window functions become more complicated in this regime, leading to biases of both signs.
We note that, in the present work, foregrounds will dominate for $k < 0.2$ and consequently this low-$k$ bias on the power spectrum is not detectable.
The proper inclusion of the window function in \herapspec\ to allow accurate estimation over all $k$ is left to future work.

\subsubsection{Foreground Models}
\label{sec:Methods:ForegroundModels}

The simulated foreground emission is constructed from two components, one of point-like sources and one of spatially smooth diffuse emission. The point source component is composed of all sources in the GLEAM catalog \citet{Hurley-Walker2017} for which a spectral model is provided with the catalog, or a number $N_{src}$ of approximately 240,000 sources, each with a power-law emission spectrum. The GLEAM catalog lacks the brightest sources, so these were added separately as point sources  according to the values in Table 2 of \citet{Hurley-Walker2017}. Fornax A was added based on the model of \citet{McKinley2015}. Explicitly for the $N_{src}$ sources each specified by their flux, spectral index and position $(F_{i}, \alpha, \vu{s}_i)$ , this component is described by:
\begin{align}
I_p(\nu, \vu{s}) & = \sum_{i=1}^{N_{src}} F_{i} \qty(\frac{\nu}{\nu_0})^{\alpha_i} \delta(1 - \vu{s} \vdot \vu{s}_i)
\end{align}
The GLEAM catalog has significant gaps in regions nominally covered by the HERA observation, notably at RA $\sim$7 hr as the Galactic anti-center transits.  
Rather than inject artificial sources, we excluded from these simulations observing times where the GLEAM catalog was significantly incomplete in the primary beam.
The diffuse emission component was simulated based on an improved version of the GSM \citep{deOliveira-Costa2008,Zheng2017,Kim2020}. In this verison of the eGSM, the spatial templates are smooth on 3 degree scales, and interpolation from the frequencies of the model maps to the desired frequency was done using a Gaussian process regression to ensure spectral smoothness.

\begin{figure}
    \centering
    \includegraphics[width=\linewidth]{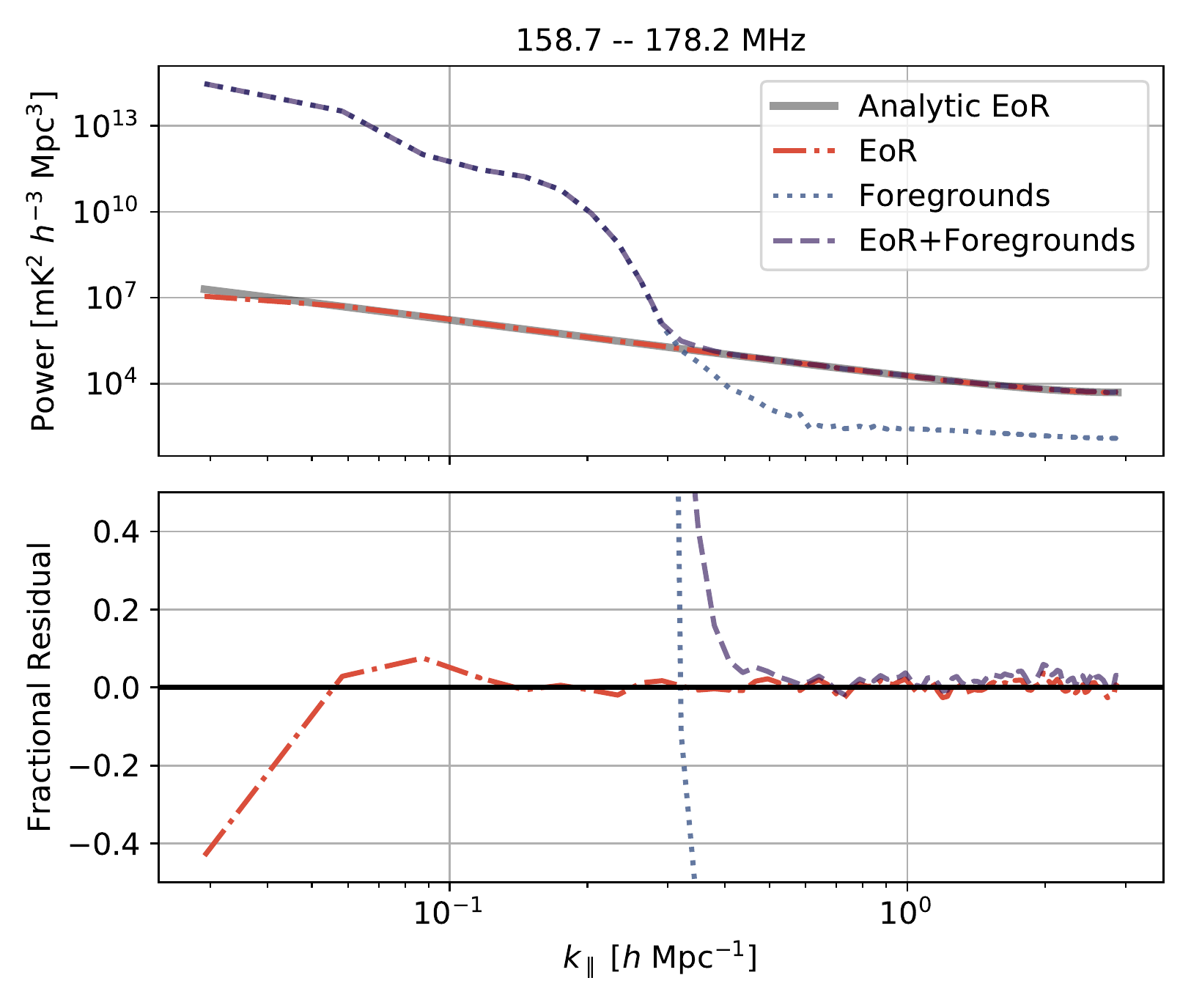}
    \caption{Noiseless power spectra estimated with \texttt{hera\_pspec}, showing power-law EoR, foregrounds (point-source plus diffuse) and their sum. Also plotted is the analytic input $P_{\rm EoR}(k)$ (corrected for aliasing, c.f. \autoref{fig:PowerSpectrumValidation}.). Lower panel shows the residuals with respect to the analytic input. Note that this figure illustrates both that the simulated foregrounds have the requisite high dynamic range (i.e. spurious spectral structure induced by the simulator is negligible) and that the power spectrum estimator correctly handles the linear sum of EoR and foregrounds in visibility space. Note that the amplitude of $P_{\rm EoR}(k)$ is lower here than in the end-to-end test (cf. \autoref{sec:EndToEnd}).}
    \label{fig:step_1pt2}
\end{figure}

The key requirements for the foreground model were that it should be representative of real foregrounds with respect to spectral smoothness and strength.
The tests in Step 1 were primarily designed to check \herapspec's ability to reproduce known input EoR power spectra in the presence of foregrounds, for $k$'s outside the foreground-dominated modes, thereby demonstrating that there are not dynamic range limitations in either the visibility simulation or the power spectrum estimation.
\autoref{fig:step_1pt2} summarizes the results of this test for the combined GLEAM and eGSM sky model.

\subsubsection{Antenna Beams}

The HERA antenna vector far-field beams $A^p_{j,\delta}(\hat{s}, \nu)$ are simulated from a detailed electrical and mechanical model using CST Microwave Studio \citep{fagnoni19}.  
Each polarized feed $p$ of an antenna $j$ responds to incident radiation from infinity in the direction $(\alpha,\delta)$ with an complex vector antenna pattern
\begin{equation}
\label{eq:BeamJones}
\vec{A}^p_{j}(\hat{s}, \nu) = A^p_{j,\delta}(\hat{s}, \nu)\hat{e}_{\delta} + A^p_{j,\alpha}(\hat{s}, \nu)\hat{e}_{\alpha}
\end{equation}
where $(\hat{\delta},\hat{\alpha})$ define an orthogonal coordinate system on the sphere, here taken to be the standard RA/Dec system.
This antenna pattern is proportional to the simulated far-field beam patterns, by the reciprocity theorem.  
The antenna patterns are assembled into a Jones matrix per antenna as
\begin{equation}
\mathcal{J}_j = \begin{bmatrix}
A^p_{j,\delta}(\hat{s}, \nu)  & A^p_{j,\alpha}(\hat{s}, \nu) \\
A^q_{j,\delta}(\hat{s}, \nu)  & A^q_{j,\alpha}(\hat{s}, \nu)
\end{bmatrix}
\end{equation}
Note that because the antenna pattern is a vector field, the appropriate representation of it in harmonic space requires spin-1 spherical harmonics.  We implemented this with custom code (see \autoref{tab:repos}).  In addition to producing a spatially smooth representation of the beam, independent of a particular pixellization, the interpolation of the spin-1 harmonic coefficients in frequency preserves the smooth frequency evolution of the beam.

\subsection{Visibility Simulation} 
\label{sec:meth:simulators}


We had several requirements for the simulation of visibilities from HERA.  
One was that our visibility simulator be able to produce  visibilities based on full-sky models of the instrument beam (including treatment of the full Jones matrix, \autoref{eq:BeamJones}) and sky (including both diffuse and point source emission).
Another was that it correctly handle drift scan visibilities, and be able to compute visibilities at the cadence of HERA time sampling over the bulk of a sidereal day, and at the full frequency resolution and bandwidth of HERA.
The resulting visibilities should do a reasonable job of reproducing the observed HERA visibilities, though we do not demand sufficient fidelity that we would be able to calibrate HERA data to the simulated visibilities.
Crucially, however, when considering the representation of the visibilities in their Fourier dual spaces (delay for frequency, fringe rate for time), the simulator should not produce numerical artifacts which adversely affect the dynamic range between bright foregrounds and regions in the EoR window.  
Specifically, with the assumption of spectrally ``smooth'' (i.e., compact in delay space) input models of the sky and beam, the simulator should not generate numerical errors that scatter foreground power to high delays.
Finally, it should be able to do these calculations in a reasonable time.

Our primary simulation engine is \rimez, an internally developed program which meets these requirements.  We describe the unique features of \rimez\ in the next section (Section \ref{sec:SimulatorMethod}).  Since any simulator will need to make approximations to allow computation in a reasonable time, and will leave out some instrument effects, we describe in Section \ref{sec:SimulatorValidation} independent checks of the simulator with  with reference calculations and make a qualitative comparison against HERA data.

\subsubsection{Simulator Method}
\label{sec:SimulatorMethod}


We take the fundamental visibility measurement equation for all four correlation products from a single baseline of a  a polarized interferometer to be
\begin{eqnarray}
\mathcal{V}_{jk} (\nu, t)
& = &  
\begin{bmatrix}
 V^{pp}_{jk} & V^{pq} _{jk}\\
 V^{qp} _{jk} & V^{qq} _{jk}\\
\end{bmatrix} \\
& = &
\label{eq:RIMEJones}
\int \mathcal{J}_j \mathcal{C}\mathcal{J}_k^{\dagger} \exp(-2\pi i \nu \vec{b}\cdot\hat{s} / c) d\hat{s}
\end{eqnarray}
where the integration is taken over the full sphere.
The coherency matrix is given by
\begin{equation}
\mathcal{C} = \begin{bmatrix}
I(\hat{s}, \nu) + Q(\hat{s}, \nu) & U(\hat{s}, \nu) - iV(\hat{s}, \nu) \\
U(\hat{s}, \nu) + iV(\hat{s}, \nu) & I(\hat{s}, \nu) - Q(\hat{s}, \nu) \\
\end{bmatrix}
\end{equation}
where $I$, $Q$, $U$, and $V$ are the images of the Stokes sky, expressed relative to the same coordinate system as \autoref{eq:BeamJones}. 
While \rimez\ is capable of fully polarized simulations, in this work we assume $Q = U = V = 0$.

In order to actually compute \autoref{eq:RIMEJones}, and in particular to address the wide field-of-view of HERA and non-trivial contribution from diffuse emission over the full sky, it is necessary to treat the integration over the sphere carefully.
\rimez\ evaluates the integral by summation over a harmonic representation of the beam, fringe, and sky terms, rather than evaluating these terms on a pixellization of the sphere, similar to the formalism in \citet{Shaw2014}. The \rimez\ implementation is based on the SSHT code for computing spherical harmonic transforms \citep{mcewen2011novel}.
Computing the visibility integral in harmonic space (for appropriate values of the maximum $\ell$) naturally handles the spherical quadrature correctly for continuous functions on the sphere, like diffuse emission and the beam.
Point sources are also included by computing their harmonic space representation; summation of the coefficients for all point sources in the simulation allows compressing an entire catalog into a single set of harmonic coefficients.
The relative orientation of the sky and the beam are handled via the $m$-mode formalism for transit telescopes \citep[e.g.][]{Shaw2014, Eastwood2019} to calculate visibilities as a function of time.  
Self-consistent auto-correlations are also produced by including the $\Vec{b}=0$ term in \autoref{eq:RIMEJones}. 

\subsubsection{Simulator Validation}
\label{sec:SimulatorValidation}



Because of the many choices inherent in implementing \autoref{eq:RIMEJones} as a numerical calculation, we independently tested \rimez\ with the goal of ensuring any systematics introduced by the simulator are below the dynamic range inherent to the \herapspec\ power spectrum estimation in the presence of foregrounds.
Undesired chromatic modulation of foregrounds $10^4$ times brighter than the background is the most challenging, but not the only, aspect to consider. 
At this dynamic range, approximations and errors usually neglected in radio interferometry become important; calculation of phases, pixellization, sky geometry, and simple coding or math errors can all be significant. 
We checked these issues first by comparison of \rimez\ calculations against analytic calculations of the visibility phase and amplitude of simple, unpolarized diffuse and point source terms in \autoref{eq:RIMEJones}.  
These tests revealed small numerical errors, but at the $10^{-10}$ level.  An additional test compared against \texttt{pyuvsim} reference simulations (Table \ref{tab:repos}), revealing differences in amplitude and phase that were primarily due to small differences in the calculation of current epoch source positions from a source catalog, which, while important for matching to data, is not relevant for the simulations here\footnote{ The complete results of this simulation comparison can be found at 
\url{https://nbviewer.jupyter.org/github/HERA-Team/hera-validation/blob/test-neg1.1.0/test-series/-1/test-neg1.1.0.ipynb}}.

To check for spectral and time smoothness, as part of Steps 0 and 1, \rimez\ was used used to generate visibilities for mock EoR and foregrounds and \herapspec\ to generate the corresponding power spectra.  For both the EoR only (\autoref{fig:PowerSpectrumValidation}) and for high-dynamic-range simulations of EoR in the presence of foregrounds (\autoref{fig:step_1pt2}), we were able to recover the input EoR, showing that the simulator was not adding additional spectral structure, as measured by the power spectrum.  
We also inspected the delay and fringe-rate transforms of the data for anomalous structure (\autoref{fig:step4_components}) and
compared simulated visibilities against real data for qualitative agreement (\autoref{fig:SimComparedToReal}).

For the test in Step 0.1, we also used \texttt{healvis}\footnote{For the purposes of long term support, \texttt{healvis}  has been incorporated into \texttt{pyuvsim}.  The standalone package is deprecated and not recommended for new projects.  }  which takes a nearly orthogonal approach to \rimez, computing \autoref{eq:RIMEJones} by pixellizing the beam and sky using \textsc{HEALpix} \citep{Gorski2005}, and performing a simple Riemann sum \citep[][Lanman et al. (2021), \textit{in prep.}]{Lanman2019,Lanman2020}.

\subsection{Noise and Instrument Systematic Simulation}
\label{sec:meth:noise}

\subsubsection{Thermal Noise}

Thermal noise is generated from a Gaussian distribution whose variance is determined on a per-time, per-frequency basis according to the amplitude of the simulated noise-free auto-correlation\footnote{The antennas are assumed to all have identical beam patterns in this work, so each antenna shares the same auto-correlation visibility prior to the application of systematic effects.} with an added receiver temperature, $T_{\rm rx}$, which is the same for each antenna, and constant with frequency.   
The standard deviation of the noise in a given (time, frequency) sample of the auto-correlation is calculated via the radiometer equation
\begin{align}
\sigma(\nu, t) &= 
\kappa(\nu) \Omega(\nu)
\frac{
\left[T_{\rm auto}(\nu, t) + T_{\rm rx}\right]}{\sqrt{\Delta\nu\Delta t}}\ \ \ {\rm Jy},
\label{eq:noise_variance}
\end{align}
where $\Delta\nu$ is the channel width, $\Delta t$ is the integration time, and $\kappa(\nu) \Omega(\nu)$ converts to Jy units (cf. \autoref{eq:KelvinToJansky}). 

Let $V_{apbq,t}$ be the visibility measured by the cross-correlation of feed $p$ on antenna $a$ with feed $q$ on antenna $b$ in time $(t, t+\Delta t)$ (we drop the implicit dependence of $\nu$ for notational clarity).
We assume the thermal noise is uncorrelated between baselines and polarizations, so we may write the noisy visibilities as
\begin{align}
V_{apbq,t}^{\rm noisy} &= V_{apbq,t}^{\rm true} + V_{apbq,t}^{\rm noise},\ \ \ a\neq b,
\end{align}
where the real and imaginary parts of $V_{apbq}^{\rm noise}$ are drawn from $\mathcal{N}(0, \sigma^2)$, with the variance given by \autoref{eq:noise_variance}.  Henceforth, we use ``true'' to denote the simulated visibility from \autoref{eq:RIMEJones}, including all sources of astrophysical emission, but excluding noise and instrumental effects except the primary beam.
The auto-correlations just have the receiver temperature added (with a signal-to-noise ratio of $\sim$1000 in the auto-correlations, this is a very good approximation); this ensures they remain real and positive definite.
The ability to reconstruct the correct power spectrum level given pure input noise was tested in Step 0.0.\footnote{Test notebook available at \url{https://github.com/HERA-Team/hera-validation/blob/master/test-series/0/test-0.0.0_noise_pspec.ipynb}}

\subsubsection{Gains}

Each antenna feed is assumed to have a direction-independent gain, representing the effects of the electronics and cables.  We assumed each antenna was represented by a diagonal Jones matrix (ignoring possible cross-coupling between the feeds). 
The average bandpass of each feed is described by a degree-6 polynomial fit to measurements of the gain derived from HERA data.\footnote{
The HERA bandpass was measured by differencing in time and frequency the cross-correlation visibilities of 19 antennas to estimate thermal noise.  
This noise was matched to a theoretical foreground and receiver temperature model to generate per-antenna, per-frequency gains. 
The resulting bandpasses for each antenna are fit by a single shared polynomial multiplied by an independent scalar amplitude per-antenna.} 
Each feed receives a unique bandpass by perturbing this average bandpass via convolution with a complex white-noise realization and subsequent application of a phase factor with a randomly generated delay and phase offset. That is, if $g_0(\nu)$ is the bandpass polynomial evaluated at the simulation frequencies $\nu$, then the antenna-based bandpass gains are given by
\begin{align}
g_{ad}(\nu) &= [g_0(\nu) \ast K_{ad}(\nu)] \exp(i2\pi\nu\tau_{ad} + i\phi_{ad}),
\end{align}
where $\ast$ indicates convolution in frequency, $K_{ap} \sim \mathcal{N}(0, 1)$ is a complex white-noise convolution kernel, and $\tau_{ap}$ and $\phi_{ap}$ are the randomly selected delay and phase offset, respectively, for antenna $a$ on day $d$ (note that the same random bandpass gain was used for each feed on an antenna).  
Note that the bandpass gains are randomized \textit{per day} rather than \textit{per time}.
This formulation ensures that the gains all vary between antennas but retain the same overall average shape.
The \texttt{hera\_sim} package was used to generate these gains.
Using these gains, we determine ``uncalibrated'' visibilities per frequency, baseline, polarization and time:
\begin{equation}
    V_{apbq,t}^{\rm uncal} = g_{a,d\ni t}g^*_{b,d\ni t} V_{apbq,t}^{\rm noisy}.
\end{equation}

Step 2 tests demonstrated that 
redundant and absolute calibration return the known input gains to machine precision, in the absence of noise.\footnote{Test notebooks available at \url{https://github.com/HERA-Team/hera-validation/blob/master/test-series/2/test-2.0.0.ipynb} and \url{https://github.com/HERA-Team/hera-validation/blob/master/test-series/2/test-2.1.0.ipynb}}
We note that because we assume the gains are band-limited in delay space, we are not testing the effect of real instrument gains which have structure beyond the smoothing scale of our {\tt smoothcal} step.

\subsubsection{Cross-coupling and Cable Reflection Systematics} 
\label{sec:meth:contaminants}

Cable reflections are captured as a per-antenna gain-like term described as
\begin{align}
\tilde{g}_a &= \prod_j^{M}\bigl(1 + A_{a,j}\exp(i2\pi\nu\tau_{a,j} + i\phi_{a,j})\bigr),
\label{eq:reflection_gains}
\end{align}
where each reflection is characterized by an amplitude ($A$), delay ($\tau$), and phase offset ($\phi$). The overall effect of reflections is the product of the $M$ different reflections (per antenna) present in the analogue signal chain.
Note that the reflection gains are generated per-antenna, and are unchanging with respect to feed and time.

Reflection gains result in the visibilities
\begin{equation}
    V^{\rm refl}_{apbq,t} = \tilde{g}_a\tilde{g}_b^* V^{\rm uncal}_{apbq,t}
\end{equation}

The cross-coupling systematic present in the H1C data \citep{Kern2020_systematics}, $V_{ab}^{cc}$, is described as
\begin{align}
\label{eq:CrossCoupling}
V_{apbq,t}^{\rm cc} &= V_{apap,t}^{\rm refl}\left(\sum_j^N A_{apbq}^{d,j}\exp(i2\pi\nu\tau_{apbq}^{d,j} + i\phi_{apbq}^{d,j})\right)_{d\ni t}
\end{align}
where each of the $N$ couplings between the auto-correlation $V_{aa}$ and the cross-correlation $V_{ab}$ is characterized as a \emph{per-baseline} reflection term. For simplicity, we only apply the cross-coupling systematic to the cross-correlations\textemdash this may be thought of as a leading-order approximation to the cross-coupling seen in the data, as cross-coupling shows up as a much smaller effect in the auto-correlations than in the cross-correlations \citep{Kern2019}.
Each of the cross-coupling parameters $A$, $\tau$ and $\phi$ are drawn randomly (see \autoref{sec:simulated_dataset} for details) per antenna $a$, feed $p$ and day $d$ (similar to bandpass gains).
Note that whether or not the cross-coupling term is {\it also} subject to reflection depends on the exact physical origin and placement in the signal chain of the cross-couplings and reflections.  In this model, the cross-coupling does {\it not} reflect.  Such a term would be second order in the (already small) coefficients, so we do not expect it to be significant relative to the current noise level.
The final ``corrupted data'' that is the input to the analysis pipeline is:
\begin{align}
\tilde{V}_{apbq,t}^{\rm corrupt} &= 
 V_{apbq,t}^{\rm refl} +  \begin{cases} V_{apbq,t}^{\rm cc} & a \neq b \\
 0 & a =b
 \end{cases}.
\end{align}
It is worth noting that we do {\it not} remove the cross-coupling systematics by fitting an equation of the form of \autoref{eq:CrossCoupling}, but rather by the method in \citet{Kern2020_systematics}.

\subsection{Flagging}
\label{sec:Methods:Flagging}

The various steps in the analysis pipeline follow very closely the description in \autoref{subsec:HERAPipeline}. We discuss in detail here the only real departure from the actual data analysis pipeline, which was our treatment of flagged data.

\begin{figure*}
    \centering

    \includegraphics[trim=35 25 35 50,clip,width=\linewidth]{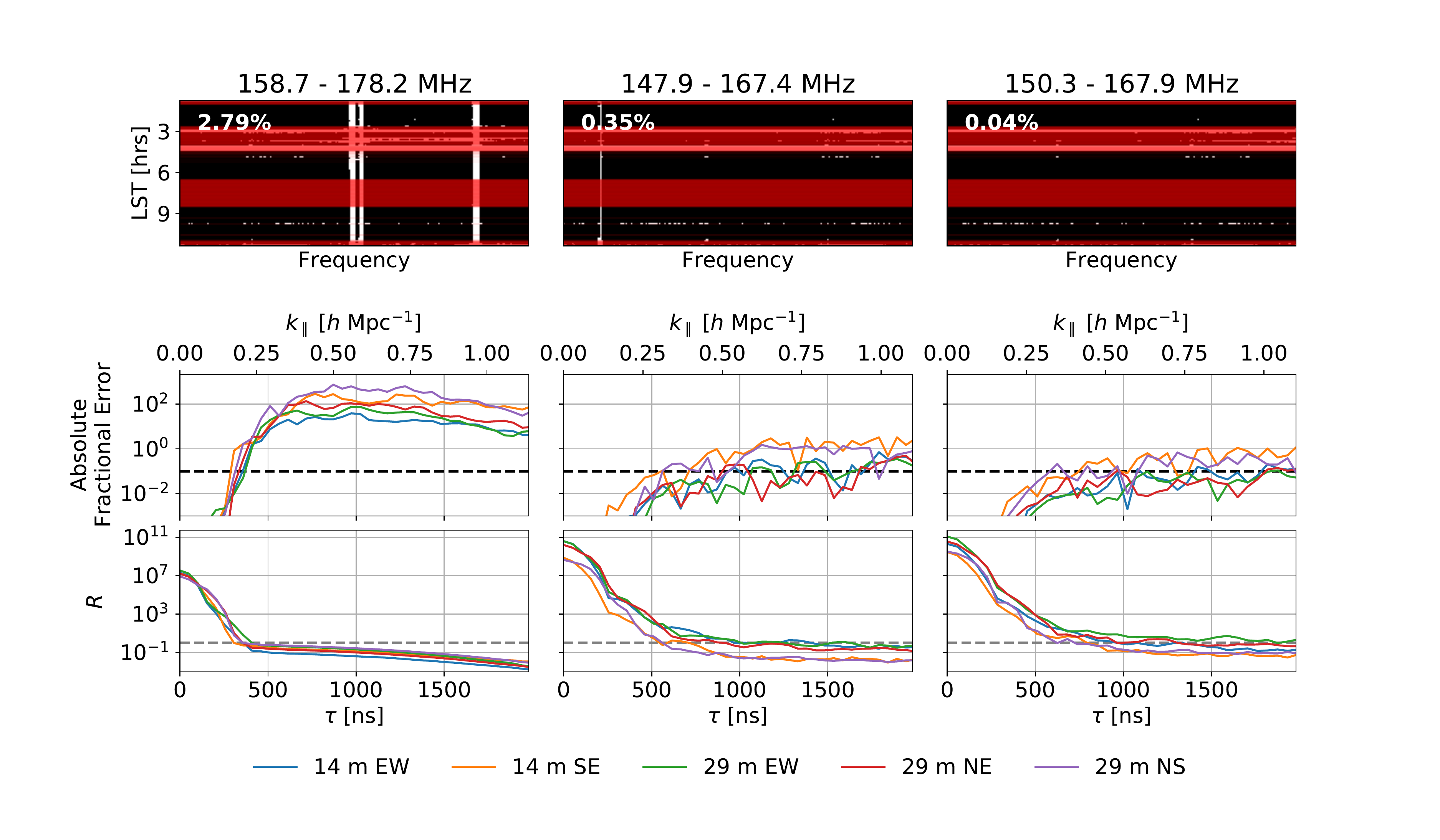}
    \caption{In-painting demonstration from Step 3.1.  Columns correspond to different spectral windows indexed by channel indices.  From left to right, the frequency channel indices corresponding to each spectral window are (600, 800), (490, 690), (515, 695).  \emph{Top row:} frequency vs. LST flagging waterfalls. White pixels represent flags in data, and red strips are integrations unused due to either flags from the in-painting process itself, flag occupancy cuts, or manual LST cuts.  The percentages in the top left corner of each waterfall represent the number of remaining flags in non-fully flagged integrations, i.e. flags within integrations not highlighted in red. \emph{Middle row:} magnitude of fractional error of recovered (in-painted) $\hat{P}(\tau)$ relative to known input as a function of delay, for various baseline types (colors).  The dashed black line marks 10\% and the delay at which the fractional error meets or exceeds 10\% is denoted as $\tau_{10\%}$.  
    \emph{Bottom row:} A metric for the dynamic range in the recovered  power spectrum, defined as 
    $R = \hat{P}(\tau) / \hat{P}(\tau_{10\%})$.  
    Good dynamic range in recovery  corresponds to higher values at $\tau=0$ ns, and greater values of $\tau_{10\%}$ (i.e. crossing the gray reference line further to the right). Note the generally increasing dynamic range and increasing $\tau_{10\%}$ with fewer flags in the center of the band, and intrinsic variability in the recovery for different baseline types.
    }
    \label{fig:Inpainting}
\end{figure*}

We chose to use the flagging patterns from the real data to test the question of how cutting data affects the power spectrum.  Recall that we did not simulate RFI or other effects that would normally be caught by the data quality portions of the pipeline and produce gaps in time and frequency.  The delay in-painting process \citep{Parsons2009,Kern2020_systematics} allows us to ``fill in the gaps'' in the frequency domain by estimating the values of the underlying Fourier modes in the delay domain.  
Since the H1C pipeline does not attempt to remove foregrounds, the bright foregrounds contribute extremely large sidelobes in delay-space if the frequency axis is Fourier-transformed with step-function-like flagging gaps. 
Filling in the gaps with an informed estimate of their true value allows us to apply Fourier techniques to (de)-flagged data, and significantly reduce these sidelobes.
However, a concern is that errors in the process will propagate power from inside the wedge into the EoR window. 
There is an additional concern that using in-painted estimates (which has no EoR signal in it) will bias the resulting power spectrum.  We show that this effect is negligible with current flagging via the end-to-end analysis.
\autoref{fig:Inpainting} shows the results of the Step 3.1 investigation of this process.\footnote{Full test notebook available at \url{https://github.com/HERA-Team/hera-validation/blob/master/test-series/3/test-3.1.0.ipynb}}

We consider a variety of flagging patterns in time and frequency taken from the data, for a variety of different baseline lengths and orientations.   We consider how accurately (in the absence of noise) the input power spectrum (foregrounds + EoR) can be reconstructed after in-painting the gappy data.  We then cast that in terms of an effective dynamic range (relative to the foreground amplitude at $\tau=0$ ns) for the reconstruction.  Generically, we find that large gaps in frequency near the center of a spectral window limit the dynamic range severely, but that for data with flagging more like the spectral windows chosen in \HOneClimit, the dynamic range exceeds $10^9$ in the power spectrum.  In the ideal case, the middle row of \autoref{fig:Inpainting} would show low fractional error in the recovered power spectrum at all $\tau$ (or $k_\parallel$).  For the choice of in-painting parameters used in Step 3.1, we chose a 10\% deviation of the recovered power spectrum from the flag free power spectrum as a fiducial marker to estimate the dynamic range, but recovery better than this would, of course, be preferred.  In the bottom row of \autoref{fig:Inpainting}, each line crosses 1 at $\tau=\tau_{10\%}$ where the recovered power spectrum deviates by 10\% in the absolute fractional error.  Better performance thus appears as a larger amplitude at $\tau=0$ ns.  Both the absolute fractional error and dynamic range are plotted in \autoref{fig:Inpainting} because they represent different aspects of our ability to recover the true power spectrum.  Small values of the absolute fractional error are required to accurately recover the power spectrum.  Large dynamic range values are required to suppress foreground contamination at higher delays.

In interpreting \autoref{fig:Inpainting}, it is important to note that the in-painting parameters are tunable.  For the work in \HOneClimit, the in-painting parameters were tuned for recovering power spectra at a dynamic range set by the expected thermal noise of \textit{observations}.  In the case of the noiseless simulated data used in Step 3.1, the in-painting parameters were tuned further, at the cost of lower computational performance, to demonstrate the ability of the in-painting process to obtain a larger dynamic range.
Thus it is difficult to translate this noise-free test directly into an impact on the measured power spectrum in the presence of noise for the full HERA analysis, and so our end-to-end simulations offer the fullest justification that the additional power added by this process at $k$'s in the EoR window averages down.
The reason we performed this test with noise-free visibilities was to isolate the accuracy of in-painting; finding the composite effects of in-painting errors interacting with thermal noise effectively requires the level of integration of an end-to-end test.

\section{End-to-End Simulations} 
\label{sec:EndToEnd}


Here we summarize the end-to-end test, which combines the independent steps discussed previously into a single run of our calibration and power spectrum pipeline.
We start by discussing the production of the simulation and systematics, and then present the results of the calibration pipeline and the power spectrum pipeline.
Lastly, we present additional tests that look carefully at the amount of possible signal loss induced by the analysis pipeline.

\subsection{The Simulated Dataset}
\label{sec:simulated_dataset}

The end-to-end simulation is not an exact replica of the dataset in \HOneClimit.  The simulated dataset contains fewer days, a narrower LST range, fewer antennas, and shorter baselines (cf. Fig. \ref{fig:array_layout}).  The LST restriction is in part due the limitations of the GLEAM sky model, which lacks sources for RA$ > 7$ in HERA's observing patch.  We show the differences between the simulated dataset and \HOneClimit\ in Table \ref{tab:SimVsReal}.  Nevertheless, the simulated dataset is sufficiently complete to capture most of the features of the real data, and to reach a comparable depth after all averaging steps were completed.

\begin{table}
    \centering
    \begin{tabular}{lll}
        \hline
        & Simulation & Data \\
        \hline
        Number of days & 10 & 18 \\
        LST range (hours) & 1.5 - 7 & 1 - 10 \\
        Number of antennas & 33 (8 flagged) & 52 (13 flagged) \\
        Total number of baselines & 300  & 741 \\
        Maximum baseline length & 84 m & 118 m \\
        \hline
    \end{tabular}
    \caption{Comparison of the parameters of the real data set and the end-to-end simulation.}
    \label{tab:SimVsReal}
\end{table}

The corrupted data was created to match the data from the H1C observing season with the computing and software resources currently available. There were three major steps in creating the corrupted data: first, we combined simulated observations of foreground emission and a reionization signal to form the true visibilities, $V^{\rm true}$; 
next, we modified the set of true visibilities to match the H1C array and observing parameters (modulo non-redundancy);\footnote{This was necessary due to the expensive original simulations having been defined for a slightly different set of antennas within the HERA array. We selected a maximal overlapping subset of the simulated baselines, interpolated to match the intrinsic H1C observation times, for use in the rest of the analysis.}
finally, the modified simulations were used to generate 10 days' worth of visibilities that were corrupted with the systematic effects outlined in \S~\ref{sec:meth:noise}-\ref{sec:meth:contaminants}. 

The base simulation consists of three components: point-source foreground emission, diffuse foreground emission, and visibilities appropriate for a power spectrum of $P_{\rm eor}(k) = 200 (k/0.2 h{\rm Mpc}^{-1})^{-2}$.
This power spectrum amplitude was chosen to produce a strong detection at $k\sim0.2\ h\ {\rm Mpc}^{-1}$, but dominated in turn by foregrounds, systematics and thermal noise at other scales.
The modified true simulations were corrupted as described in \S~\ref{sec:meth:contaminants}. 
Here we specify the values used in the data corruption for the end-to-end test.
Recall that $V_{ab}^{\rm uncal}$ has two parameters per antenna: the delay, $\tau_a$, and the phase offset, $\phi_a$. We drew delays randomly from a uniform distribution spanning (-20, +20)\,ns, chosen via manually tuning to matched observed features in H1C data, and the phase offsets from a uniform distribution on [0, 2$\pi$). The bandpass gains used for this work vary only between antennas, nights, and as a function of frequency; we did not add any LST variability to the gains. 

\begin{figure}
    \centering
    \includegraphics[width=\linewidth]{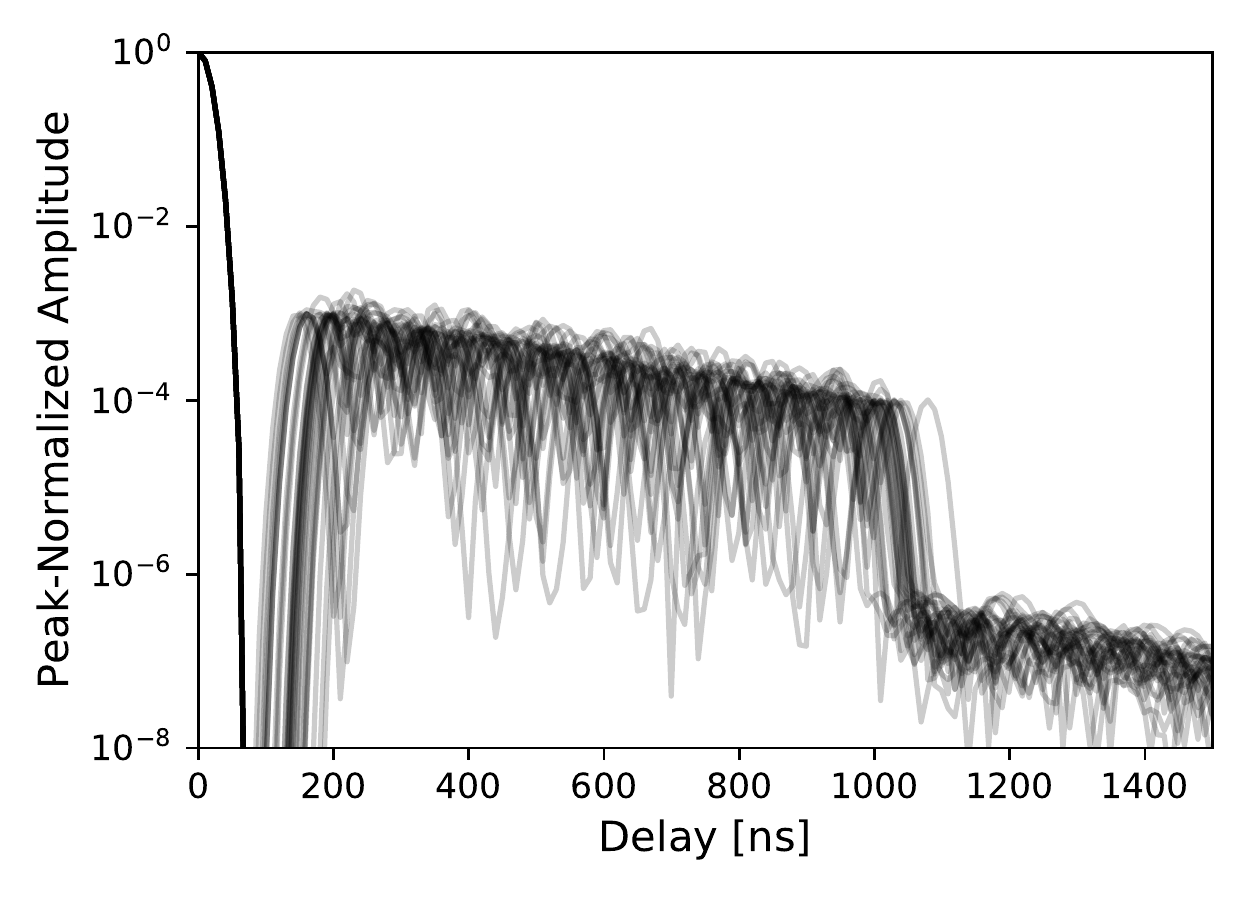}
    \caption{Peak-normalized delay spectra of simulated reflections. Each line is a different antenna. Note the relatively narrow spread in the reflection at 200 ns contrasted with the relatively large spread in the reflection at 1200 ns. Low-level features beyond 1200 ns are second-order reflection terms.}
    \label{fig:reflection_gains}
\end{figure}

Recall that the effect of a single cable reflection is characterized by an amplitude, delay, and phase offset (\autoref{eq:reflection_gains}).
As implemented, the reflections were split into two categories: single cable reflections and a reflection ``shoulder'' meant to model a series of sub-reflections. The former consisted of two relatively high-amplitude reflections with random per-antenna delays of $(200 + 10\epsilon_{200})$\,ns and $(1200+30\epsilon_{1200})$\,ns, with relative gain amplitudes of $3\times10^{-3}$ and $8\times10^{-4}$ respectively, and $\epsilon$ a standard normal variable.
These delays and relative amplitudes were chosen to match the observed systematics in the H1C system \citep{Kern2020_systematics}.
The sub-reflection shoulder consisted of 20 individual reflections uniformly located  between a delay range of 200 -- 1000\,ns, with amplitudes following a power law in delay from $10^{-3}$ to $10^{-4}$.
Each reflection's delay is perturbed by a Gaussian offset with a scale of 30\,ns, and their amplitudes were perturbed randomly by $1\%$ of their assigned value.
All reflection terms have their phases drawn from a uniform distribution from $[0, 2\pi)$ and are not varied across frequency, time and observing night.
An example of the reflection gains in delay space is shown in \autoref{fig:reflection_gains}.

The cross coupling model is simulated in a similar manner as the reflection shoulder, meant to match systematics observed in the H1C system.
We generated an independent set of reflection terms (amplitude, delay, phase) for each baseline, per night. We characterize the cross-coupling with ten reflections (each determined by an amplitude, delay and phase). The ten delays were spaced linearly between 900 -- 1300\,ns.
Amplitudes are regular in log-space, $A = 10^{-(\tau-100)/200}$ (going from $10^{-4}$ at $\tau=900$\,ns to $10^{-6}$ at $\tau=1300$\,ns), but offset by a random normal variable with scale 0.01\% of the amplitude.
Phase was drawn from a uniform distribution from $[0, 2\pi)$.
Cross coupling is simulated as described by \autoref{eq:CrossCoupling}, where the systematic is the product of the auto-correlation visibility with each of the reflection coupling terms.
The phase of the reflections are similarly drawn randomly for each reflection and antenna from a uniform distribution from [0, $2\pi$).

\begin{figure*}
    \centering
    \includegraphics[width=\linewidth]{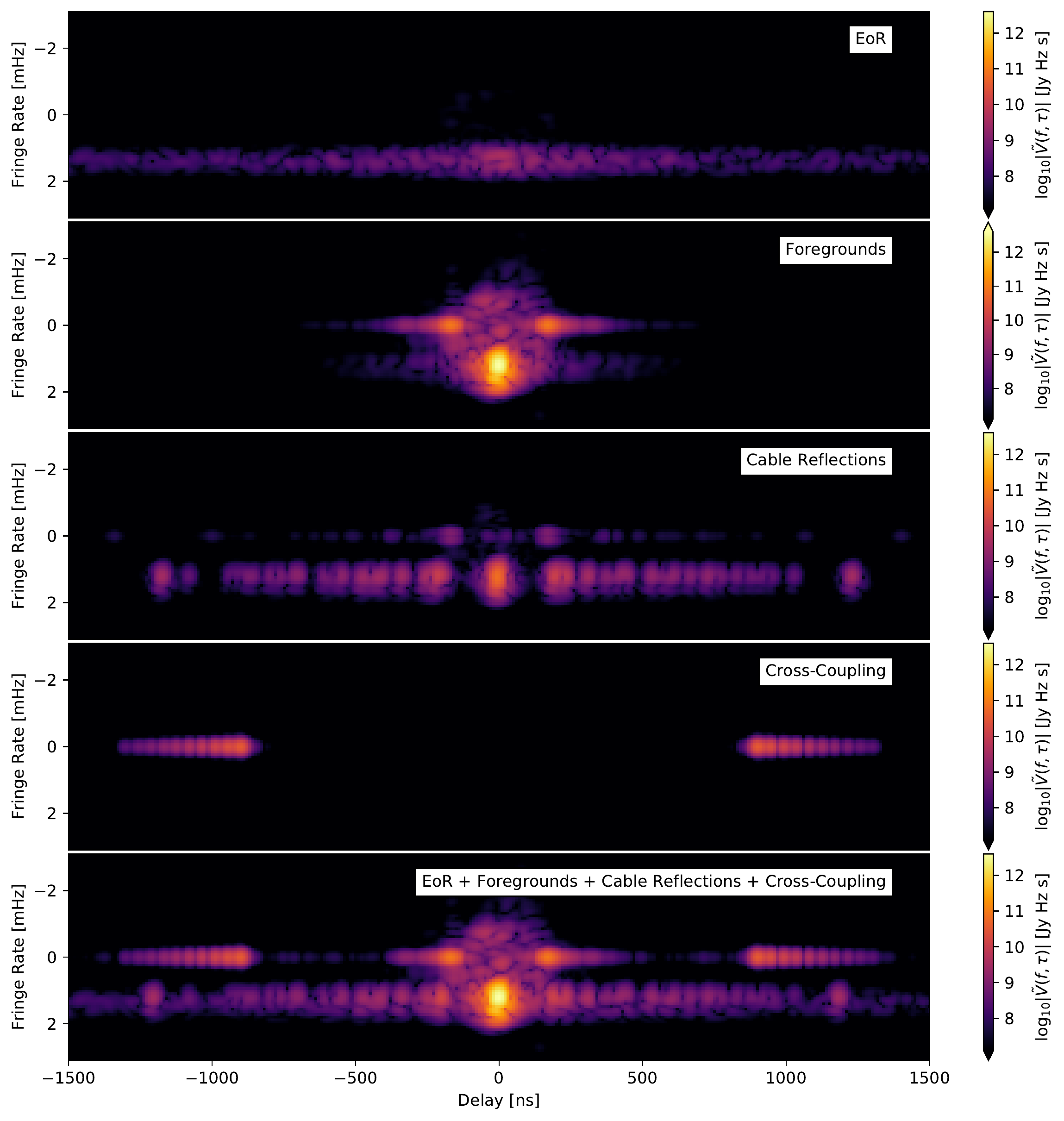}
    \caption{Delineation of different simulation components for a 51-meter WNW-oriented baseline (44-meter westward East-West projection, 25-meter northward North-South projection), Fourier transformed along the frequency and time and frequency axes (x- and y-axis respectively). The fringe-rate/delay basis is useful for highlighting unique physical characteristics of each component.}
    \label{fig:step4_components}
\end{figure*}

Figure \ref{fig:step4_components} shows the various components described above in fringe-rate / delay space, for a 51-meter WNW-oriented baseline.
The top panel shows the EoR component, demonstrating that it primarily populates positive fringe-rates (as it is a statistically isotropic and sky-locked signal, and the baseline has negative EW component).
The foreground component has a positive fringe-rate component as well, but also demonstrates significant power at fringe-rates near zero, peaking at a delay corresponding to the length of the baseline.  The ``pitchfork effect'' due to the monopole described by \citep{thyagarajan15a} also shows excess power at the delay corresponding to the baseline length; the observed effect here is clearly related to this when resolved in both fringe-rate and delay.
Importantly, the fringe-rates occupied are dependent on the EW component of the baseline, which is why cross-coupling cannot be removed from baselines with a small EW component without incurring signal loss.
The cable reflection plot shows the convolution of the reflection terms with the foreground signal, showing that it acts to smear the foreground signal horizontally across delay.
Lastly, the cross-coupling signal is fairly independent of the other terms, and occupies near-zero fringe-rate modes.

\begin{figure*}
    \centering
    \includegraphics[width=0.9\linewidth]{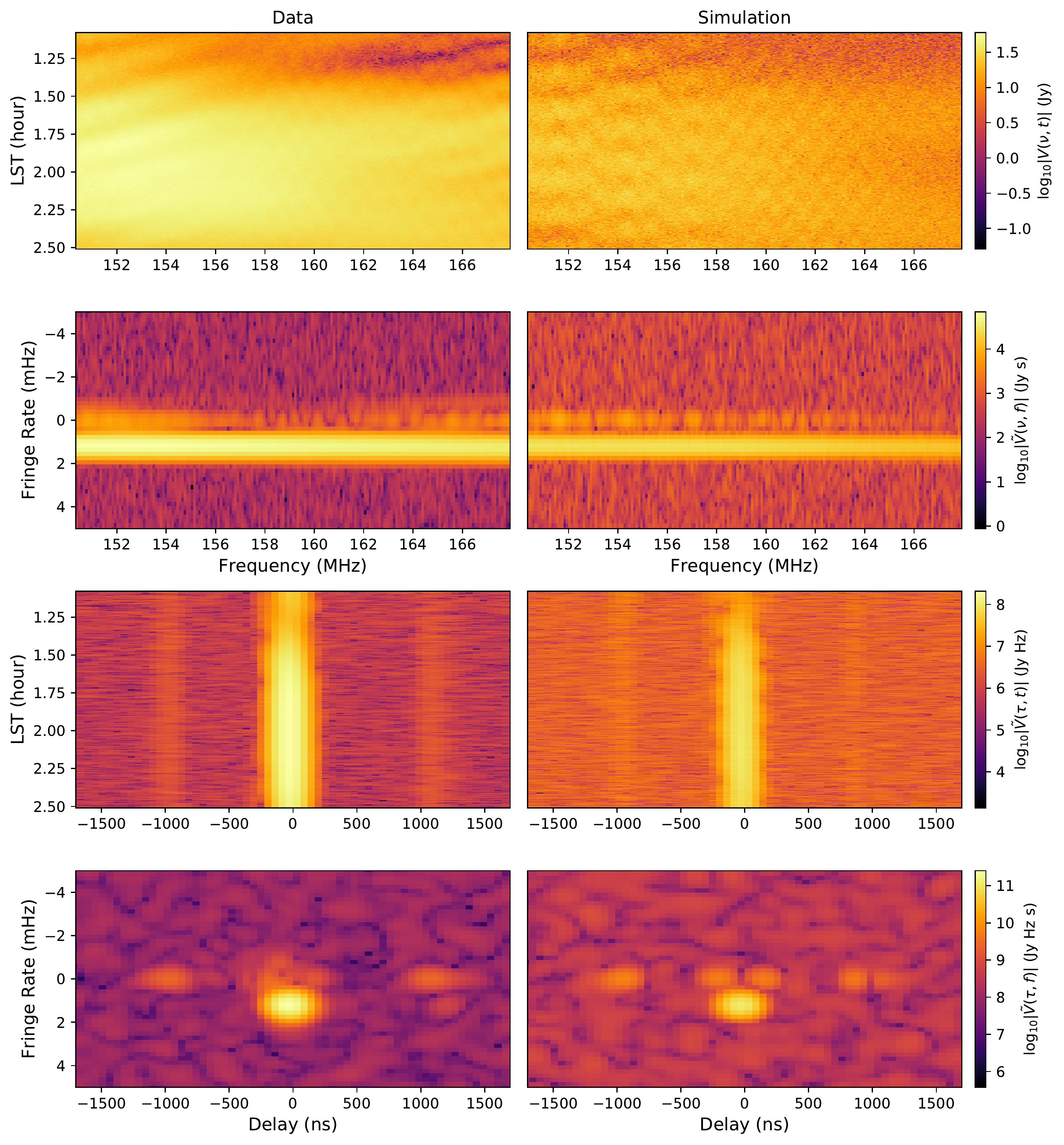}
    \caption{Comparison of simulated data (right) with observation data (left) for the same LST range, spectral window, and baseline. Each set of plots shows the four possible choices of Fourier transformed versions of the data. The observation data has been redundantly and absolutely calibrated, but not treated for cable reflections or the cross-coupling systematic. The observation data has also been LST-binned, so it has a substantially lower noise level than the simulation. While the simulation and observation data look strikingly similar, there is a clear difference in the qualities of the high-delay systematic: the simulated version is much more symmetric in delay, and it appears to be somewhat brighter than it is in the observed data. 
    }
    \label{fig:SimComparedToReal}
\end{figure*}

In Figure \ref{fig:SimComparedToReal}, we show a comparison of the simulated data to real data, highlighting the simulation's ability to capture the general features observed in the data.

\subsection{Calibration Results}
\label{sec:step4_analysis_results}

\begin{figure}
    \centering
    \includegraphics[width=\linewidth]{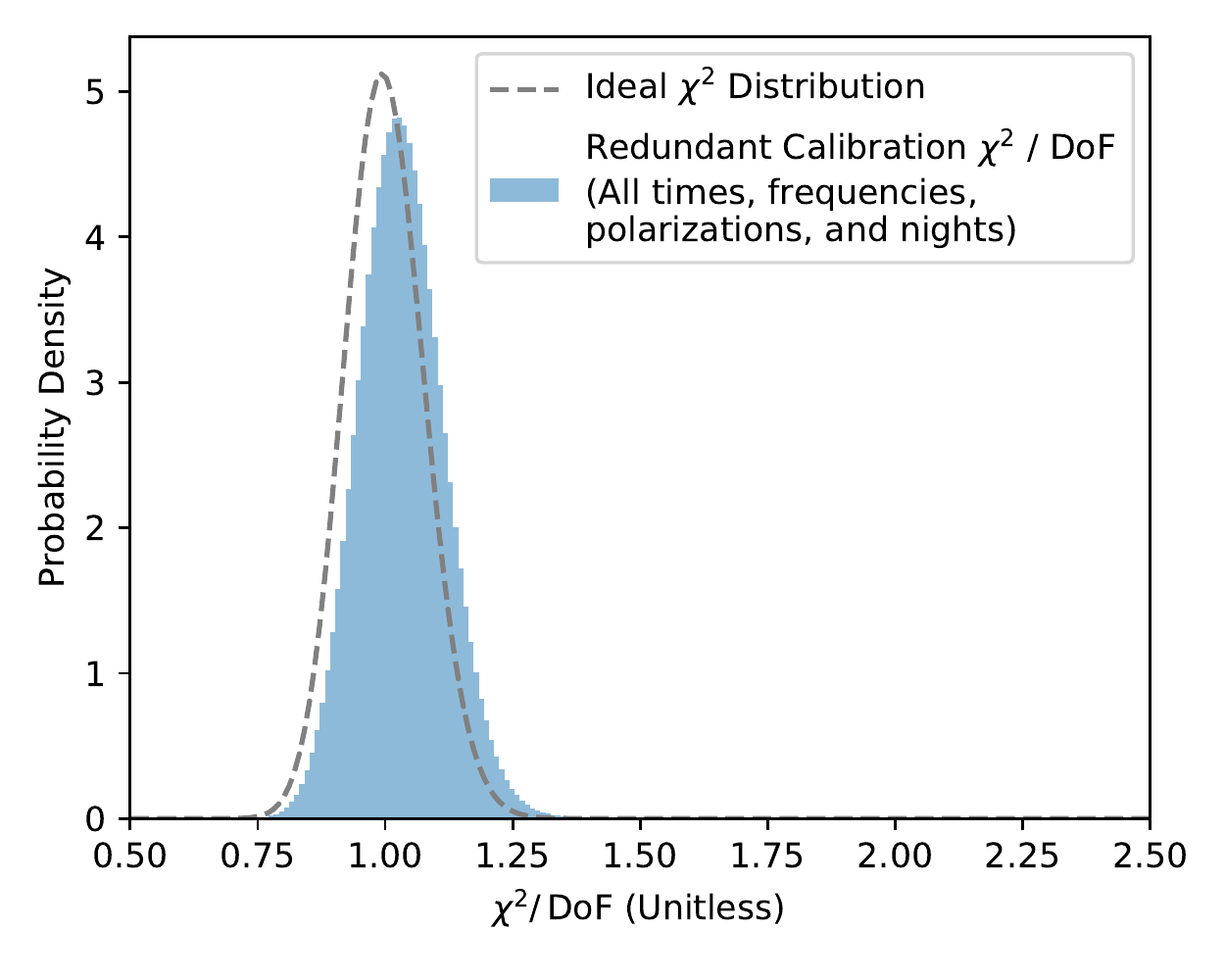}
    \caption{The success of redundant-baseline calibration can be assessed by examining the difference between raw visibilities and the visibility model for by assuming redundancy and antenna-based gains. This is quantified by $\chi^2$ per degree of freedom, which was defined in \citep{Dillon2020}, and can be compared to a theoretical expectation (in this case with $\text{DoF} = 164$). Here we can see that the simulated distribution of $\chi^2$ nearly matches the expected distribution; we expect a mean value of $\chi^2$ / DoF of 1, we observe 1.03. This is substantially better the observed distribution, which peaks around 1.3 to 1.4 \citep{Dillon2020}. The key difference is that while both the validation simulation and real data contain baseline-dependent cross-talk systematics (an additive effect which breaks the assumption of redundancy), the simulation does not contain any antenna position errors or antenna-to-antenna variation of the primary beam which likely account for most of the observed deviation from 1.}
    \label{fig:chisq_dist}
\end{figure}

Each night of the corrupted simulated data are first passed through the H1C calibration pipeline, which employs a direction-independent calibration of the XX and YY polarizations for each 10.7 second time integration.
The performance of redundant calibration on the H1C system is described in \citet{Dillon2020}, and can be summarized by the final reduced $\chi^2$ of the gain solutions.
\autoref{fig:chisq_dist} shows the reduced $\chi^2$ (blue) of the estimated gains from the simulated data set compared to an idealized, pure-noise distribution (dashed-grey), showing a slight positive bias indicative of excess variance due to the presence of baseline-dependent cross-coupling systematics that break the redundancy condition.

\begin{figure}
    \centering
    \includegraphics[width=\linewidth]{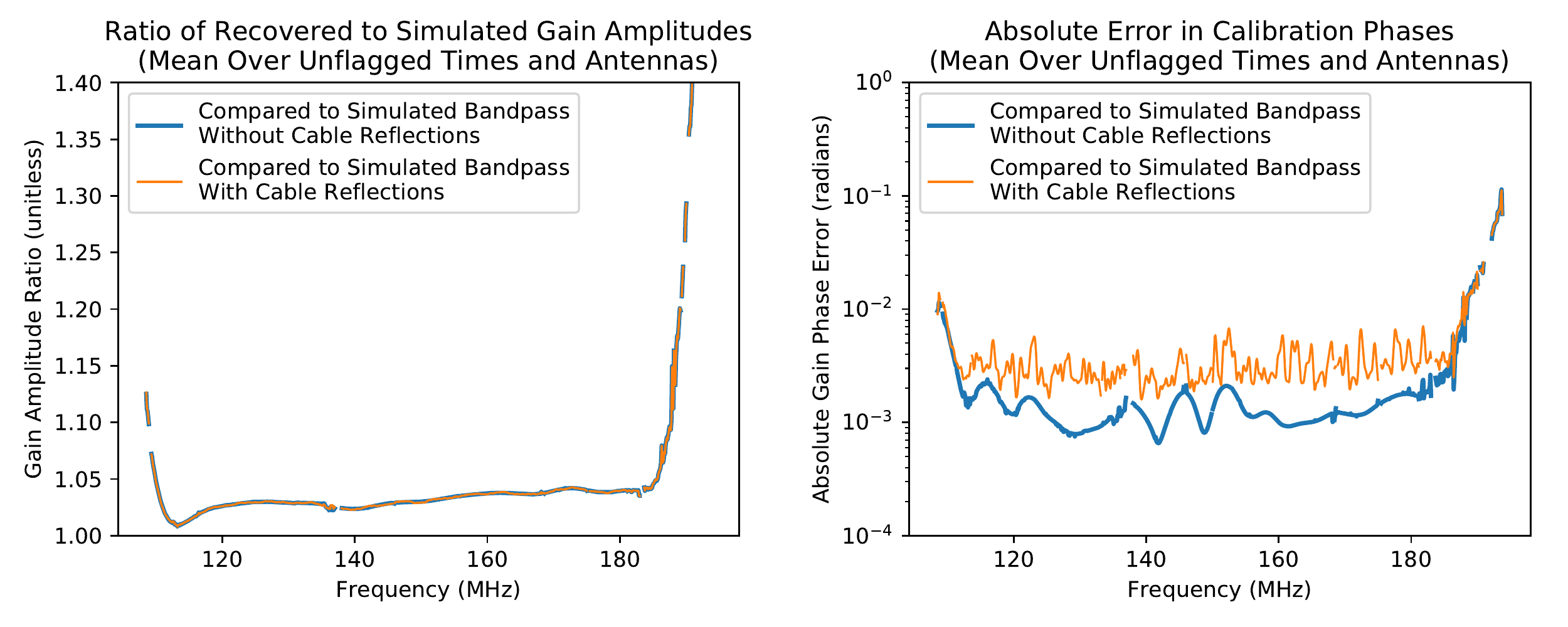}
    \caption{After the calibration pipeline, the inferred gain solutions are quite close to the true simulation gains, with amplitude errors at the few-percent level and phase errors at the few-milliradian level. Calibration errors are likely due to a range of factors including thermal noise, cross-talk systematics, and smoothing of features in the inferred calibration solution at the 100\,ns delay scale. The latter explains both the spectral structure in the phase errors, which is largely contained to higher harmonics, and the increased errors at the band edges, since the smoothing was performed with a Tukey window ($\alpha = 0.3$) which down-weights discrepancies at high and low frequencies. The cable reflections, which dominate the true gains at high delay, are intentionally smoothed out of the calibration solutions and corrected only after LST-binning. The smoothing out of real spectral structure from cable reflections produces the dominant phase calibration error at most frequencies, but is subdominant to the few-percent-level amplitude bias seen in the left panel, which is due to a small bias in absolute calibration (see Figure \ref{fig:abscal_bias}). While this level of gain error and bias should be factored into a final power spectrum and errors, it is unlikely to produce substantial signal loss from decoherence.}
    \label{fig:calibration_errors}
\end{figure}

\begin{figure}
    \centering
    \includegraphics[width=\linewidth]{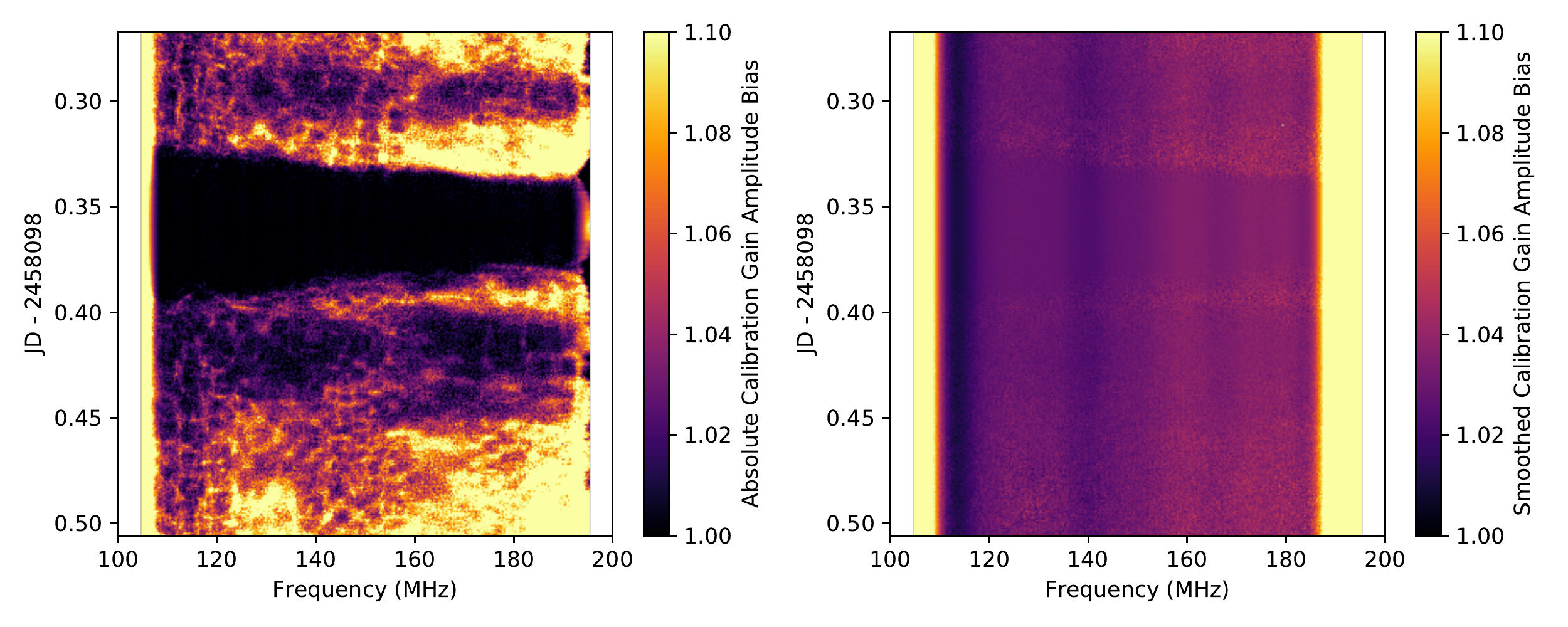}
    \caption{Here we show the bias in our amplitude calibration of a single simulated night (JD 2458098), averaged over all unflagged antennas. Because we are calibrating the overall amplitude of a noisy data set with a lower-noise set of ``model'' visibilities using a logarithmic linearization of the calibration equation, the time and frequency bins with low SNR return gain amplitudes that are biased high \citep{Boonstra2003}. The opposite is observed at the transit of Fornax A around JD 2458098.36, which produces very high SNR visibilities and suppresses the bias, leading to near perfect gain amplitude recovery. The right panel shows the gain bias after frequency and time smoothing of the gain. We see that the bias is now effectively time-independent, but has a slight dependence on frequency, which is accounted for when forming power spectra over different parts of the band.}
    \label{fig:abscal_bias}
\end{figure}

When we compare the estimated gains to the true gains used to corrupt the data, we find that the gain phases are recovered to good precision, while the recovered gain amplitudes are biased slightly high (\autoref{fig:calibration_errors}).
Gains biased slightly high will result in the calibrated data being biased slightly low.
This bias comes from a time-varying signal-to-noise ratio (SNR) of the visibilities and our choice of absolute calibration technique (cf. \autoref{fig:abscal_bias}).
Calibrating noisy data with a low SNR can lead to biased estimates of the gain amplitude when using a logarithm to linearize the antenna-based calibration equation \citep{Boonstra2003}, as we employ here. Comparing the gain solutions to the known simulated gains we find solutions biased high by roughly $4\%$ (left panel of \autoref{fig:calibration_errors}).

This absolute flux scale bias can also be well-quantified by imaging the true model visibilities and the post-calibrated data and comparing the fluxes of a bright point sources near beam-center.
This reveals an amplitude biased low by $\sim8\%$ and varying slowly with frequency.\footnote{In the real H1C data set, this bias is closer to $4\%$, which is due to the larger range in right ascension included in the real H1C data compared to the validation data set.}

To correct this, we multiply all estimated power spectra and their $1\sigma$ errorbars by the measured bias of 1.11 for the low band and 1.15 for the mid band (\autoref{tab:signal_loss}).

Other sources of uncertainty from absolute calibration can also impact the overall error budget.
These effects include: i) the overall uncertainty on the flux scale at $\sim10\%$ \citep{Hurley-Walker2017} and ii) the change in the gain amplitude due to ambient temperature drift during a nightly observation, which is not corrected for in the H1C pipeline and is estimated to be roughly $5\%$ \citep{Kern2020_abscal}.
While these sources of uncertainty are added in quadrature with the errorbars in \HOneClimit, we do not do so here as these effects have not been included in simulation.

\begin{figure}
    \centering
    \includegraphics[width=\linewidth]{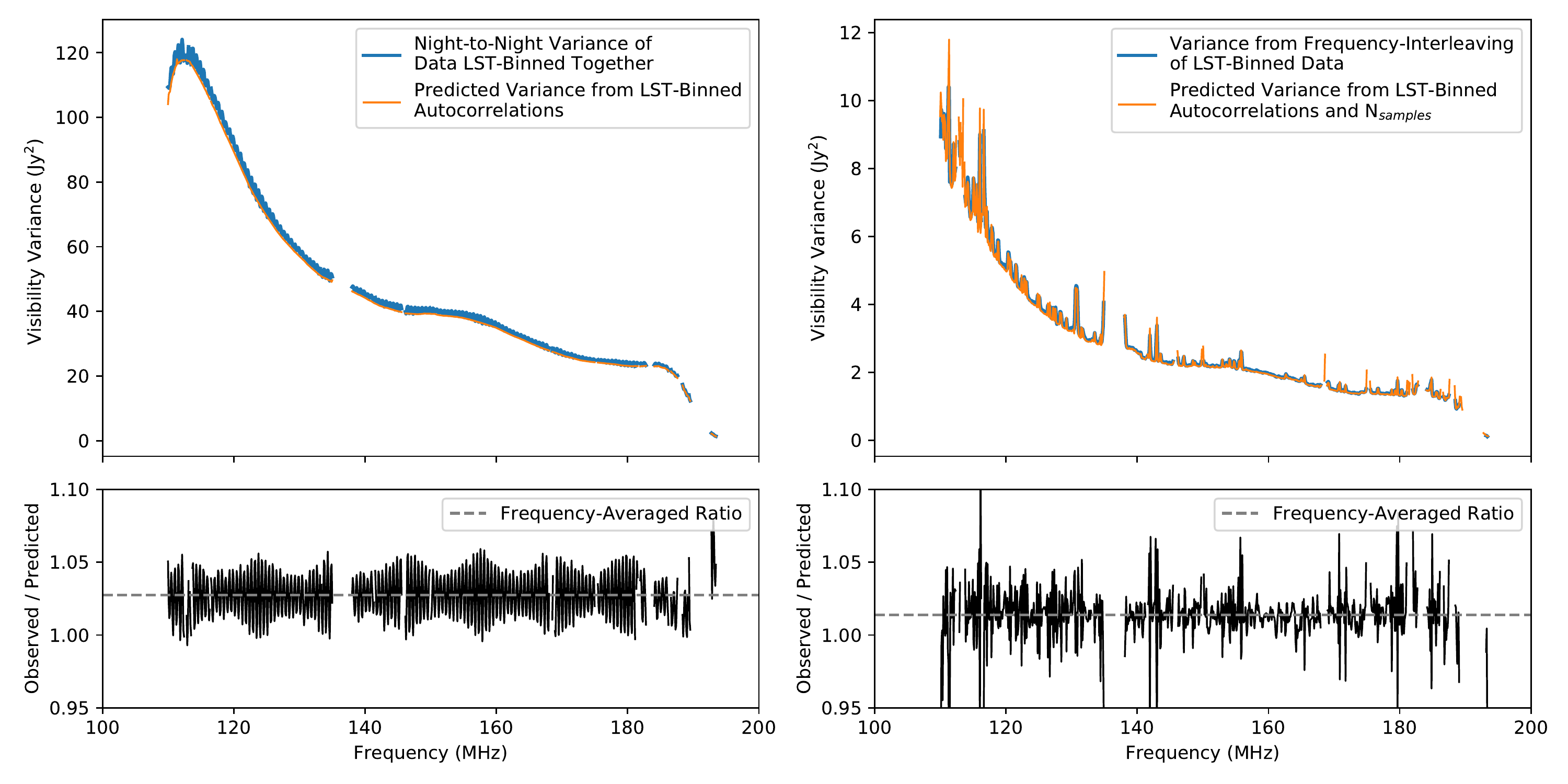}
    \caption{Confirmation that noise in LST-binned visibilities matches  expectations. \emph{Left panel:} night-to-night variance over 10 nights (averaged over unflagged baselines and times in the LST range of 6.464 -- 6.817 hours) compared against noise predicted by LST-binned autocorrelations (see \cref{eq:noise_variance}). \emph{Right panel:} variance calculated from the same data using frequency differencing compared to normalized predicted noise from autocorrelations and $N_{\rm samples}$. In both panels we drop we drop any time or frequencies with $N_\text{samples} < 10$ before averaging in order to account for RFI gaps. Both metrics indicate a close match with the predictions.
    }
    \label{fig:LST_variance}
\end{figure}

The model visibilities used in absolute calibration are not constructed in an identical manner as \HOneClimit, which used a CASA-based pipeline to calibrate a few independent fields, then stitched them together to construct a set of model visibilities for all LSTs while finally low-pass filtering the visibilities across frequency to reduce noise and fill-in gaps due to RFI.
Instead, here we simply take the sum of the foreground, EoR and noise visibilities as our representative model visibilities, sidestepping the question of calibration uncertainties due to an incomplete sky model for the time being.
However, as demonstrated in \citet{Kern2020_abscal} and \citet{Dillon2020}, the gain smoothing procedure applied to the post redcal + abscal gains is meant to filter-off any fast time and frequency structure in the gains that might be generated by such issues.

Each night is calibrated independently and then binned onto a uniform grid in LST and coherently averaged together (known as LST binning).
We show via two methods in \cref{fig:LST_variance} that to within a few percent, the noise in our LST-binned visibilities matches our expectations. 
One way to estimate the noise in LST-binned visibilities is simply to measure the variance of all visibilities that are to be binned together after re-phasing to a common phase center. In our case, since our LST-binned data set has a cadence of 21.4 seconds (twice as long as the nightly simulations with a 10.7 second cadence), we simply compute the variance of all (up to) 20 visibilities in a single LST bin from the 10 nights simulated. Since the LST binner is also estimating the mean visibility, we use the unbiased estimator of the variance (i.e. we use Bessel's correction). 
The left panel of \cref{fig:LST_variance} shows that this matches closely the noise we expect on each input visibility, as inferred from the calibrated autocorrelations using \cref{eq:noise_variance}. 
Another way of estimating the noise in the LST-binned visibilities is to take the frequency-interleaved difference---in this case, $V_{ij}(\nu) - \frac{1}{2} V_{ij}(\nu - \Delta \nu) - \frac{1}{2} V_{ij}(\nu + \Delta \nu)$. This gives an estimate of the noise variance at frequency $\nu$, though discontinuities in $N_\text{samples}$ complicates it slightly. Regardless, the right panel of \cref{fig:LST_variance} shows that the observed noise again matches the expectation for how the noise in the visibilities should integrate down quite well, accounting for the number of samples. In both panels we drop any time or frequencies with $N_\text{samples} < 10$ before averaging in order to account for RFI gaps.

After binning, the averaged visibilities are passed through systematics treatment \citep{Kern2019, Kern2020_systematics}.
This involves modeling the smooth foregrounds and in-painting the model in the remaining RFI flags in the data (\autoref{sec:Methods:Flagging}).
Next the auto-correlations are used to model antenna-based reflections in the signal chain.
A total of 28 signal chain reflection terms are iteratively solved for, chosen by visual inspection of the residuals, and the algorithm is only provided the rough location in delay space where we expect reflections to appear (150 - 1500 ns).
After calibrating out the reflection terms, we apply the \citet{Kern2019} procedure for modeling and subtracting off the slowly time-varying cross-coupling systematics.  
Note that this cannot be done reliably for baselines with a projected East-West length less than 14 meters without substantial signal loss \citep{Kern2019}, so we flag all baselines that do not meet this requirement.

Next, the visibilities are coherently averaged in time with a 214 second averaging window, having first phased the different time integrations to a common pointing center.
Lastly, the instrumental XX and YY visibility polarizations are summed to construct a pseudo-Stokes I visibility as $V_I = (V_{XX}+V_{YY})/2$.
Recall that many of these analysis steps are tested individually (\autoref{fig:validation_matrix}), but here we present the effects of these steps on the fully integrated power spectrum.

\subsection{Power Spectrum Recovery}
\label{sec:step4_pspec_results}

Power spectra are formed in the same manner as described in \HOneClimit.
To summarize, we form delay spectra \citep{Parsons2012_dspec} in two spectral windows, which we refer to as Band 1 and Band 2, spanning 117 -- 132 MHz and 150 -- 168 MHz, respectively.
Note that because we apply a Blackman-Harris apodization function across each spectral window, their equivalent bandwidth is half of the full bandwidth.
Power spectra are formed by cross multiplying every pair of non-identical baselines within a redundant set.
We then calculate two sets of errorbars: a theoretical noise RMS given the measured system temperature ($P_N$), and a semi-empirical errorbar that accounts for the signal-and-noise cross terms in the power spectrum, $\tilde{P}_{\rm SN}$ \citep[cf. Table 3 of][]{tan2020}.
All incoherent averaging (i.e. averaging after forming the power spectra) is weighted inversely by $P_N^2$, but the final quoted errorbars come from $\tilde{P}_{\rm SN}$.
The justification for using these errorbars is outlined in the discussion (Section 5) of \citet{tan2020}.
\HOneClimit\ outline three broad LST ranges (or ``fields'') that are used for forming an averaged power spectrum.
With the slightly smaller LST coverage studied in this work, we look at two similar fields, spanning 1.5 -- 2.8 hours LST, and 4.4 -- 6.4 hours LST.
Power spectra are formed from three datasets: the full data without systematics treatment, the full data with systematics treatment, and just the EoR component of the data.

\begin{figure}
    \centering
    \includegraphics[width=\linewidth]{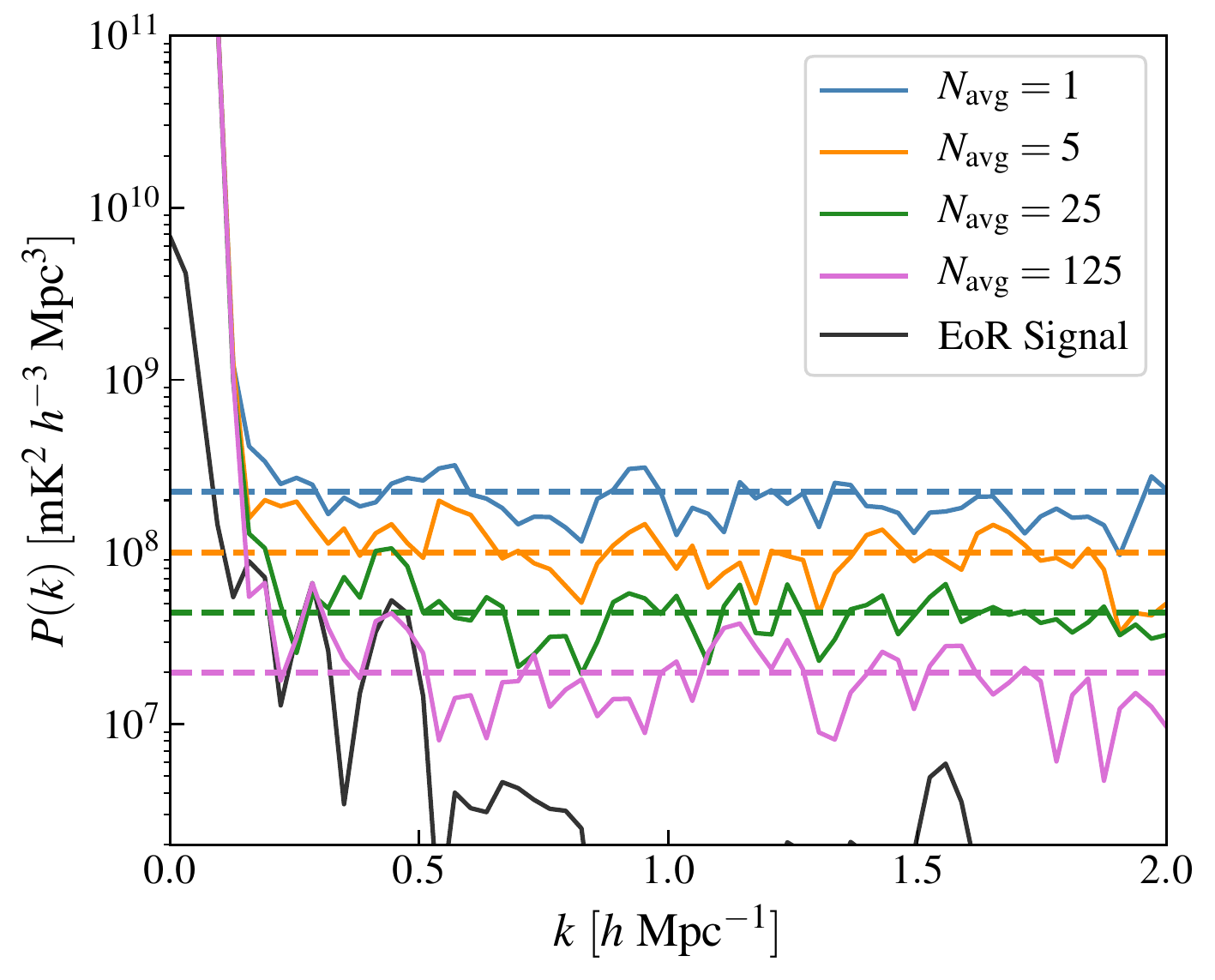}
    \caption{Power spectra from Band 1 after successive incoherent redundant baseline averages. In each case we plot the averaged power spectra (solid) and their corresponding $P_N$ given the amount of averaging (dashed), which marks the noise amplitude if the power spectra are noise dominated. In each case we see good agreement between the two for $k > 0.2\ h\ {\rm Mpc}^{-1}$, except for the final case where at low $k$ we begin to detect a signal; note the black line which is the fully-integrated spectrum for this baseline type.}
    \label{fig:test4_noise_integration}
\end{figure}

The first check on our power spectra is to ensure that our noise estimates agree with the data at different stages of integration.
This has recently been studied and validated for HERA simulations and real data \citep{Kern2020_systematics, tan2020}, but we repeat the exercise here for completeness.
\autoref{fig:test4_noise_integration} demonstrates the impact of incoherent averaging within a single redundant set.
We show successive averages of redundant baselines with an increasing number of baselines in each average.
We also plot the propagated thermal noise uncertainty $P_N$ (dashed), which agrees well with the power spectra outside of the foreground dominated region for $k > 0.2\ h\ {\rm Mpc}^{-1}$.
Note that the final average (magenta) shows an increase in power at low $k$, which is foreshadowing a low-level detection of the EoR signal in the simulated data.

\begin{figure*}
    \centering
    \includegraphics[width=\linewidth]{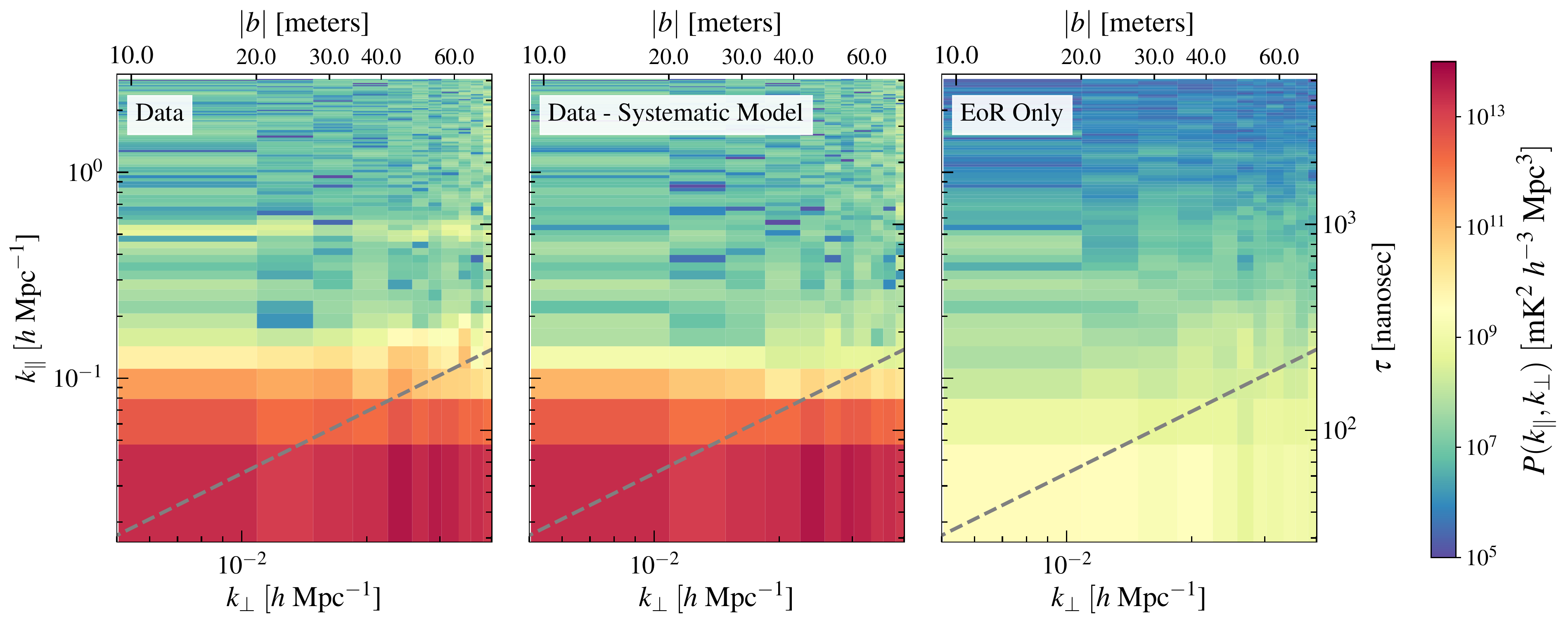}
    \caption{Two-dimensional delay spectra for the end-to-end test (step 4). The left panel shows the result having skipped the systematic subtraction step (and thus contains all of the extra instrumental systematics injected into the data), the center panel shows the full end-to-end run with systematic subtraction, and the right panel shows the EoR only dataset. The grey dashed line marks the foreground horizon (i.e. the wedge).}
    \label{fig:test4_wedge}
\end{figure*}

After averaging all baseline-pairs within a redundant group we average the remaining time bins within each LST range, leading to a cylindrically averaged $P(k_\parallel, k_\perp)$ power spectrum per field per spectral window.
\autoref{fig:test4_wedge} shows this for Band 1 of the first LST range.
The left panel shows the full dataset, the middle panel shows the data after systematic treatment, and the right panel shows the EoR-only dataset.
In all panels, the grey dashed line shows the extent of the foreground wedge from the baseline horizon.
We  expect some amount of leakage beyond this line simply due to the sidelobes of the apodization function applied before taking the visibility Fourier transform.
What is apparent from \autoref{fig:test4_wedge} are the systematics at $k_\parallel\sim0.5\ h\ {\rm Mpc}^{-1}$ that are effectively suppressed by the modeling and removal step.
This systematics treatment also suppresses power just beyond the foreground wedge.
As discussed in \citep{Kern2020_systematics}, this is a result of the fact that foregrounds entering from the horizon (and thus lying near the wedge border in $k_\parallel,k_\perp$ space) are slowly time-variable and are therefore partially filtered off with the cross-coupling filter.
So although \autoref{fig:test4_wedge} becomes less wedge-like with the application of the cross-coupling filter, this is due to filtering which impacts the edge of the wedge. The wedge will manifest on longer baselines \citep[e.g.][]{Kern2020_systematics}.

\begin{figure*}
    \centering
    \includegraphics[width=\linewidth]{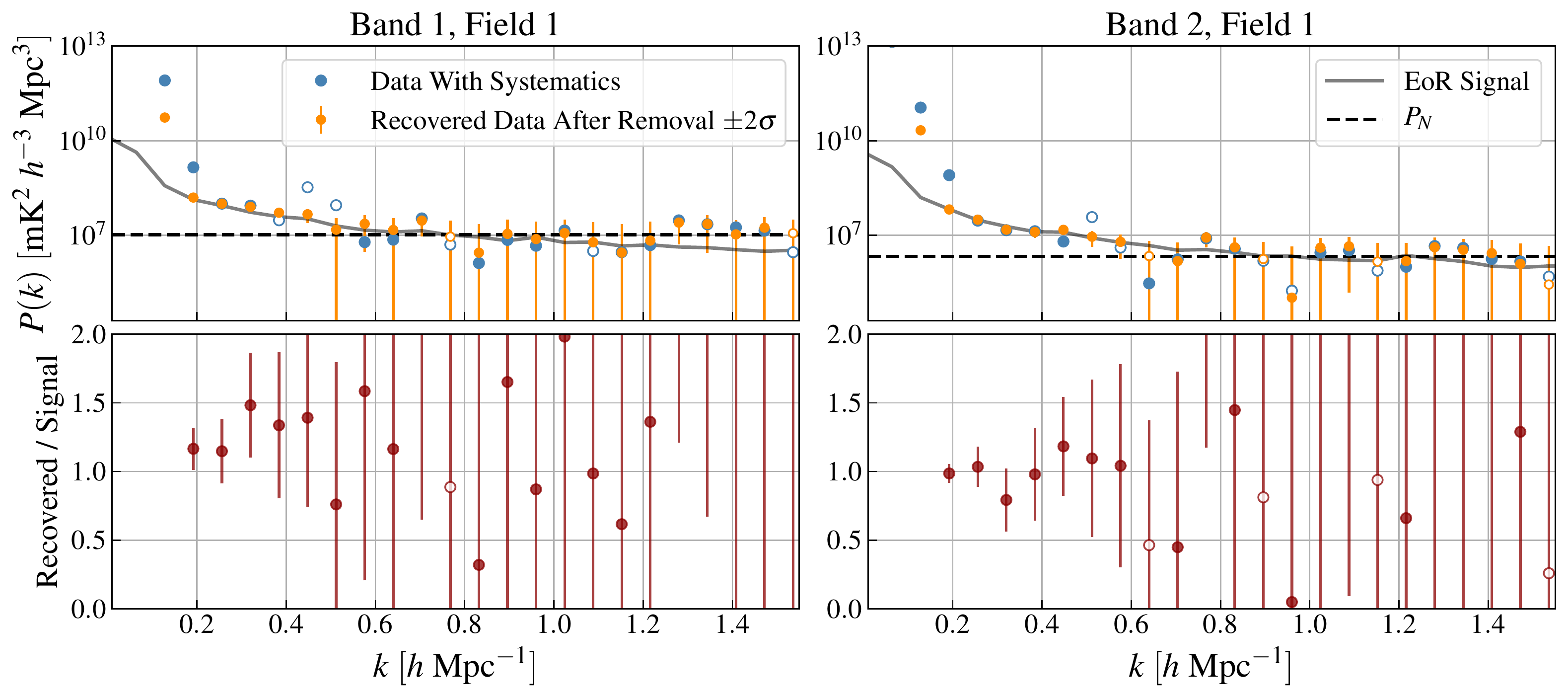}
    \includegraphics[width=\linewidth]{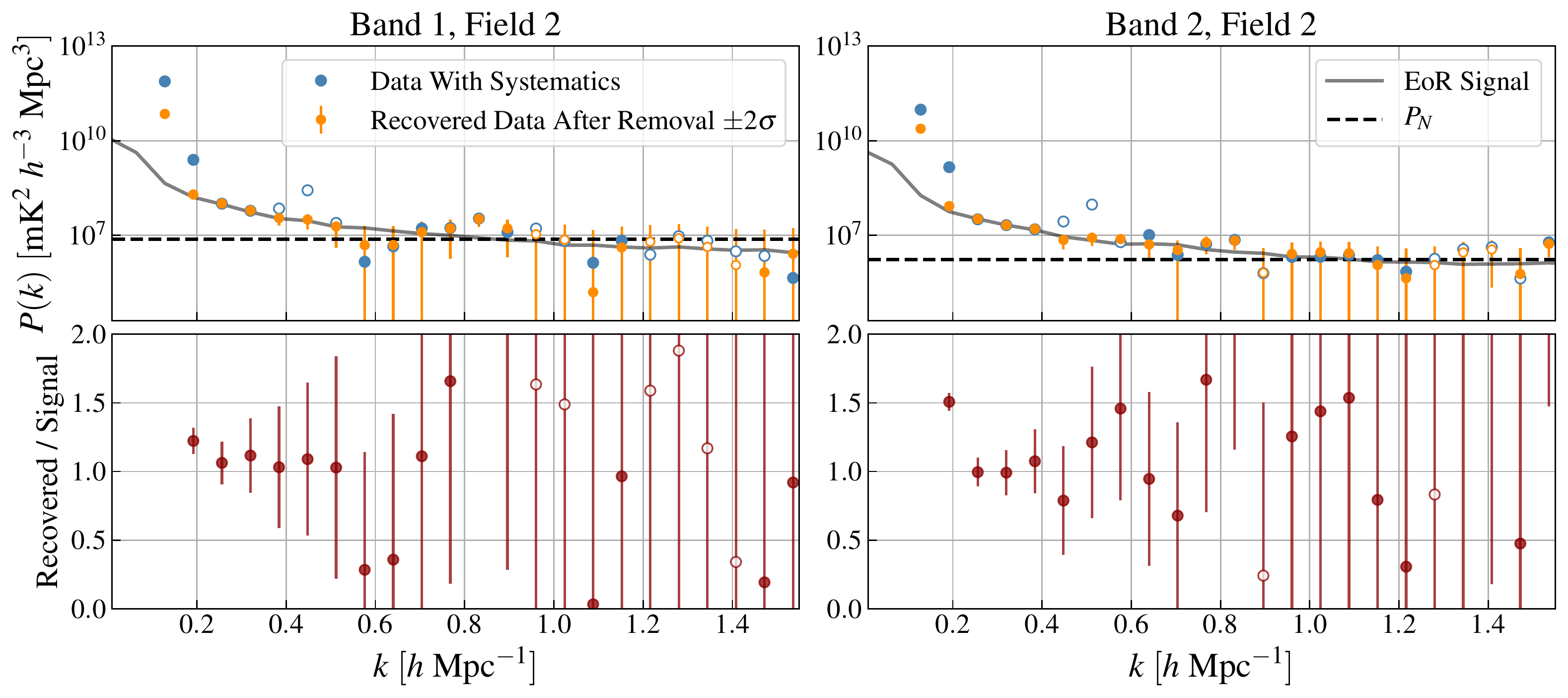}
    \caption{Recovered power spectra of the end-to-end test (step 4). We plot the data before systematic treatment (blue), after systematic treatment (orange) with its $2\sigma$ errorbars, as well as the intrinsic EoR signal (grey) and the noise floor (black dashed). The top panel shows the first LST cut (1.5 -- 2.8 hours) and the bottom panel the second LST cut (4.4 -- 6.4 hours). The subpanels plot the ratio of the recovered (orange) over the EoR (grey), showing unbiased recovery of the intrinsic EoR signal at all $k$ to within the estimated errorbars.}
    \label{fig:test4_lstcuts}
\end{figure*}

Lastly, we group the cylindrical power spectra in bins of constant $|k|$ and spherically average them to get our final 1D power spectra.
\autoref{fig:test4_lstcuts} shows these results for the first LST cut (top panels) and the second LST cut (bottom panels), for both Band 1 (left panels) and Band 2 (right panels).
We plot the data before systematic treatment (blue points), after systematic treatment (orange) points with $2\sigma$ errorbars, as well as the EoR-only dataset (grey).
Open circles denote negative band powers, which are plotted as positive for visual clarity.
The sub-panels show the data after systematic treatment divided by the EoR-only dataset, with the $2\sigma$ errorbars overlaid.
Recall that the amplitude of the EoR signal was chosen specifically to allow for a detection of the signal at low $k$, with its significance decreasing at higher $k$.
The salient points we draw from \autoref{fig:test4_lstcuts} are: 1) the EoR signal is recovered to within the errorbars across all $k$ modes\footnote{For the second LST cut there is an outlier at $k\sim0.85\ h\ {\rm Mpc}^{-1}$ for both Band 1 and Band 2. It is odd that these outliers occur at the same Fourier mode, although an outlier or two is not entirely unexpected as the errorbars plotted are $\pm2\sigma$. At the very least, it is a high outlier, so concerns of signal loss are not an issue. More work is needed to understand if this is a statistical or systematic outlier.}, and 2) the systematics at $k\sim0.45\ h\ {\rm Mpc}^{-1}$ are suppressed down to the measured EoR amplitude.
Importantly, the recovered power spectra (orange) match the EoR signal at low $k$ where we detect the signal at $\sim10$ times the noise floor, and at high $k$ the power spectra are consistent with noise.

\begin{figure}
    \centering
    \includegraphics[width=\linewidth]{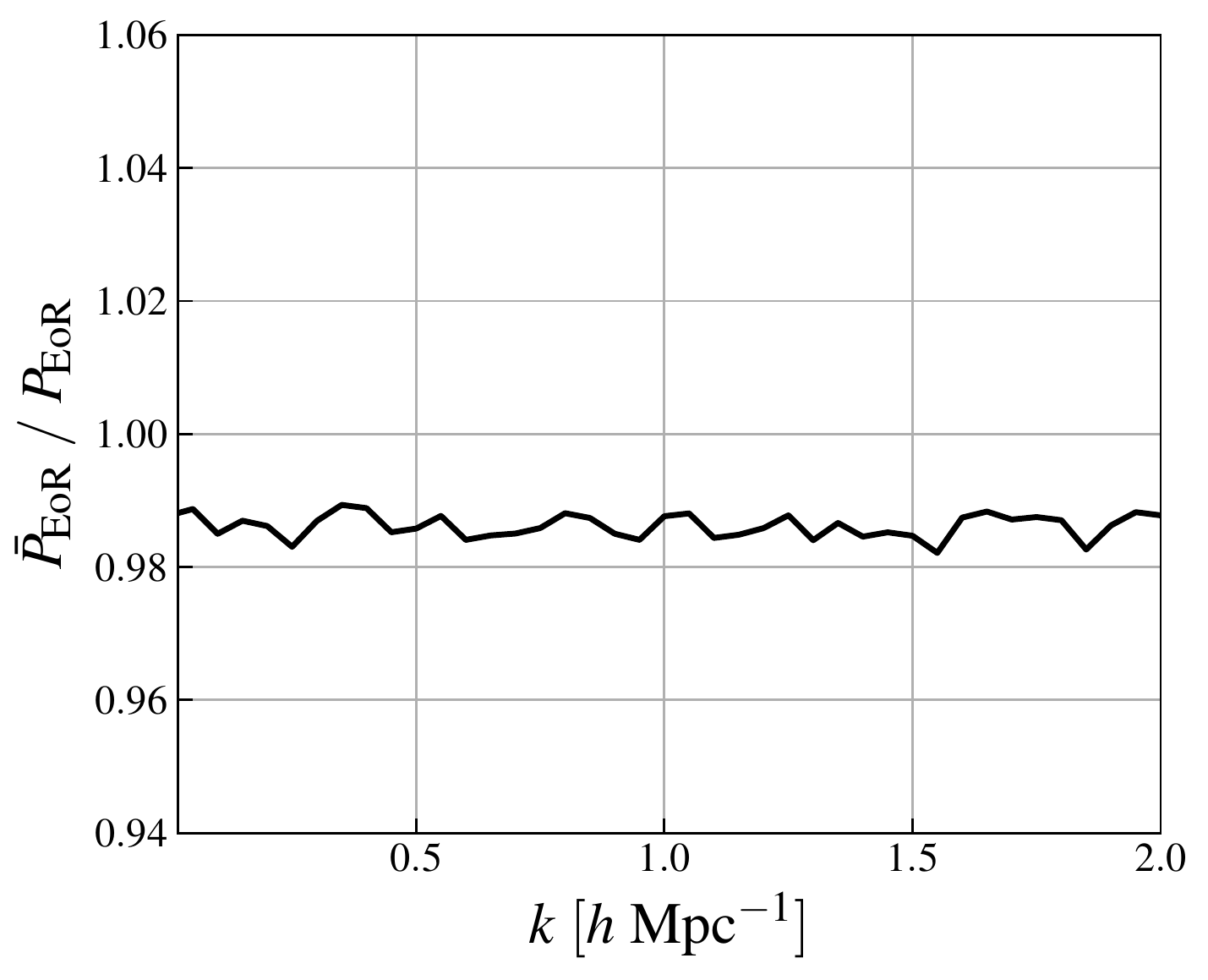}
    \caption{Signal loss test for the LST averaging step in the H1C analysis pipeline. This measures the amount of signal loss induced by coherent averaging of the visibilities across LST. The numerator of the ratio is the power spectrum of the EoR-only visibilities having averaged over a 7-minute window, while the denominator is the same data product with a time averaging window of only 43 seconds. This step induces $\sim1\%$ signal loss, which is deemed negligible compared to other limiting uncertainties. This result has been verified against different visibility simulators and different EoR models \citep{Kern2019}.}
    \label{fig:test4_avg_EoR}
\end{figure}

Using the data products at hand, we also perform additional tests targeted at particularly sensitive components of our analysis pipeline. 
One analysis step not quantified in the unbiased recovery seen in \autoref{fig:test4_lstcuts} is the amount of loss induced by coherent time averaging (i.e. LST averaging or fringe-rate filtering).
Over the course of a drift-scan observation, one can coherently average different time integrations that are closely spaced together relative to the overall beam crossing time after re-phasing them to a common pointing center.
Using Monte Carlo simulations of an ensemble set of mock, $P(k)\propto k^0$ EoR observations, \HOneClimit\ claim that they can coherently average their visibilities over a 528 second window and only induce $\sim1\%$ signal loss in the measured EoR power.
We use the data products in this work (which recall use a $P(k)\propto k^{-2}$ EoR model) to confirm that this specification is met.
\autoref{fig:test4_avg_EoR} shows this test, comparing the ratio of the EoR-only power spectra having first averaged the visibilities over a 528 second window over the power spectra with a 43 second averaging window.
We show that, as expected, this induces $\sim1\%$ loss in power that is constant across Fourier $k$ modes.

\begin{figure}
    \centering
    \includegraphics[width=\linewidth]{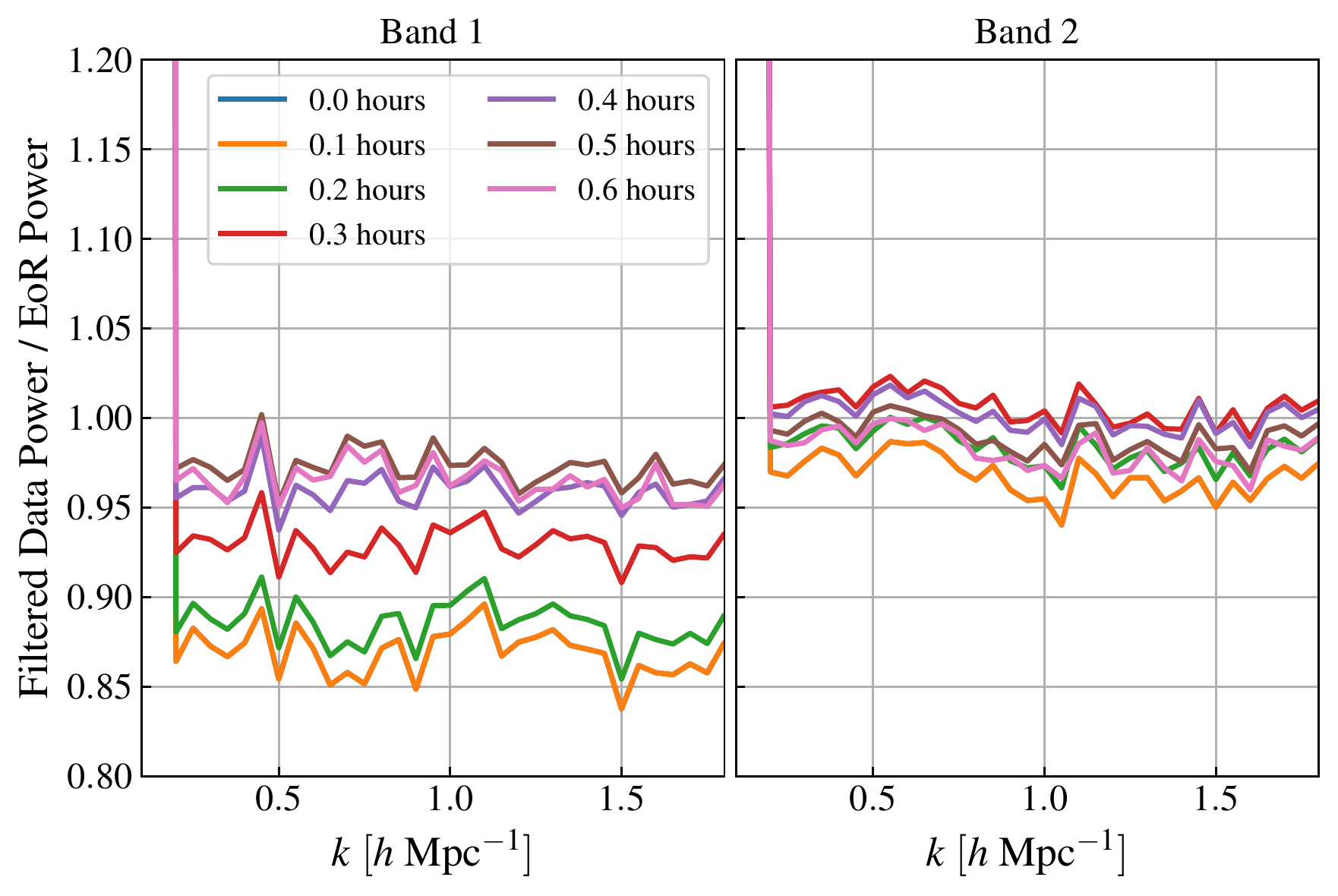}
    \caption{A measurement of the amount of loss induced by the cross-coupling high-pass time filter as a function of how much of the data near the edges of our time axis we flag. Because of edge-effects, the cross-coupling filter leads to more loss for time bins near the bounds of our time axis. We show here that by flagging 30 minutes on either side, we minimize this loss, with a residual loss of $3\%$ ($1\%$) in power for Band 1 (Band 2).}
    \label{fig:xtalk_filter}
\end{figure}

Another somewhat sensitive step in our analysis chain is the filtering of cross-coupling systematics, which is performed by applying a high-pass filter across the time axis.
Recall that such a filter is designed to reject the slowly variable systematics, while retaining the vast majority of the EoR sky signal \citep{Parsons2016, Kern2019, Kern2020_systematics}.
One complication to this is the impact of the time-edges when working with finitely sampled data.
Near the time edges the properties of the sharp Fourier filter are degraded, and in our case we observe slightly more loss than the original specification \citep{Kern2019}.
We can mitigate this effect by flagging the time bins near the edges after filtering.
\autoref{fig:xtalk_filter} shows a demonstration of this on a dataset that contains only the EoR signal and a cross-coupling systematic.
We remove the systematic in the same way, but now in averaging the power spectra we flag all time bins within some buffer of the time edge, showing that the amount of loss converges with a 30 minute buffer.
In the end, this results in a residual $3\%$ ($1\%$) scale-independent loss in power after cross-coupling subtraction for Band 1 (Band 2).

All of the steps discussed in this work which were found to lead to loss are summarized in \autoref{tab:signal_loss}, most of which are on the order of a percent.
The bias discussed in \autoref{sec:step4_analysis_results} is not technically signal loss in the traditional sense, but is still a bias that results in an under-reporting of the EoR signal, therefore we include it in this table.
Note that \HOneClimit\ also explore the impact of coherent baseline averaging within a redundant group, which can in principle lead to signal loss.
We do not explore this currently as baseline non-redundancy is not within the intended scope of this work, but future work will incorporate this aspect into the validation pipeline presented here.

\begin{figure}
    \centering
    \includegraphics[width=\linewidth]{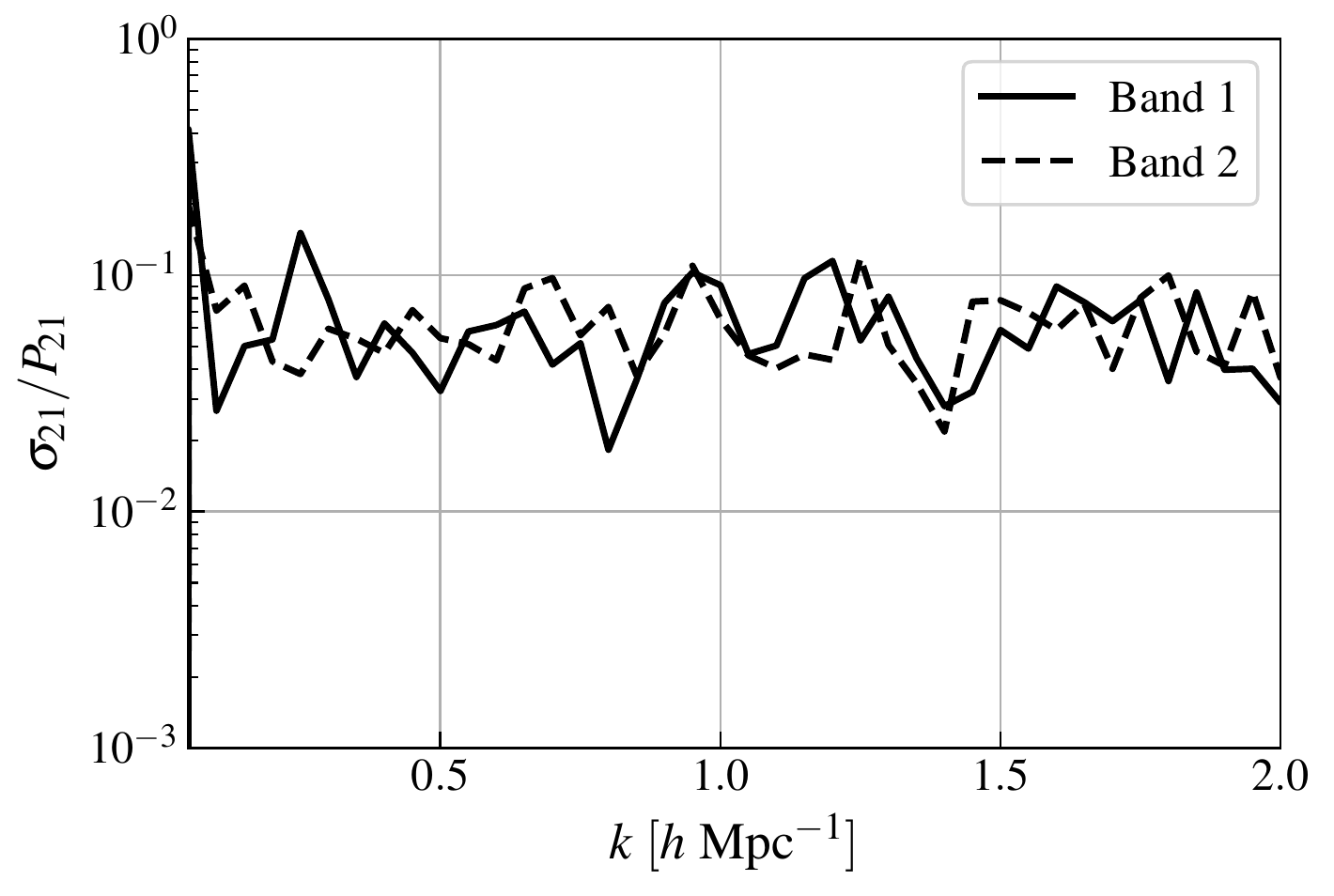}
    \caption{Fractional ($1\sigma$) cosmic variance uncertainty on the power spectrum for two hours of drift-scan observations with HERA-37 is on average $\sim5.5\%$. This is computed by taking the RMS of the EoR-only power spectra across the two hour time range, and then dividing by the effective degrees of freedom set by the HERA beam crossing time of one hour.}
    \label{fig:eor_cosmic_variance}
\end{figure}

Lastly, another metric we can pin down with our validation simulations is the expected level of cosmic variance on the EoR signal after all of our coherent and incoherent averaging.
\citet{Lanman2019} quantify this in a similar manner using Monte Carlo simulations of a mock EoR field, and find that for a HERA-37 spherical power spectrum averaged over 8 hours LST the cosmic variance ($1\sigma$) peaks at around $2\%$ of fractional power.
Pushing our own EoR simulation through the analysis pipeline discussed in this work and averaging the power spectra over the second LST range (2 hours spanning 4.4 -- 6.4 hours LST), we find a fractional ($1\sigma$) cosmic variance uncertainty of $\sim5.5\%$ for both Band 1 and Band 2 (\autoref{fig:eor_cosmic_variance}).
As discussed in \citet{tan2020}, this is currently a subdominant contributor to the total error budget.

\begin{deluxetable}{lc} 
\tabletypesize{\footnotesize} 
\tablewidth{\linewidth} 
\tablecaption{
Systematic Loss in Analysis
\label{tab:signal_loss}
}
\tablehead{Analysis Step & Fractional Bias}
\startdata
Absolute Calibration & -11\% (-15\%) \\ [.1cm]
Cross-Coupling Filtering & -3\% (-1\%) \\[.1cm]
LST Time Averaging & -1\% (-1\%) \\[.1cm]
\enddata 
\tablecomments{Percentage loss in power for Band 1 (Band 2), which are corrected for after forming the power spectrum and are constant for all $k$ modes. Redundant baseline averaging is also explored in HC20 as a possible source of percent-level loss, but is not studied in this work.}
\end{deluxetable}

\subsection{Blind Test with Parallel Pipeline}

To double-check that our primary power-spectrum analysis pipeline indeed induces minimal signal loss and makes a clear detection of the input 21\,cm power spectrum, we performed a blind analysis of the mock data with an alternate power spectrum estimator\footnote{Use of the word blind here might not be preferred by blind people, however the  term is used pervasively in science to describe a technical procedure which is not well served by synonyms. We keep the term for now to avoid confusion as the practice is introduced to the field of 21\,cm cosmology.}.

Parallel, or \emph{shadow}, analysis is a powerful validation technique that has been adopted for several published results \citep{Jacobs2016,Barry2019a,Trott2020}. 
These analyses are expensive in both researcher and computer time, and thus are often limited
in the amount of data processed in parallel, and may share common pre-processing steps.
Nevertheless, they provide some measure of confidence that the reported result is not unique to a particular analysis.
Errors made in the absence of such testing have commonly been associated with power spectrum estimation \cite{Paciga2013,Cheng2018}, so this is where we choose to focus our efforts.
We perform a parallel power spectrum analysis of the calibrated and LST averaged simulation product using the \textsc{simpleDS} pipeline \citep{Kolopanis2019},
verifying that it also reproduces the expected result.

Our shadow analysis followed the procedure described in \citep{Kolopanis2019}. 
Power spectra were formed by cross-multiplying redundant baselines and errors estimated by calculating the expected sensitivity according to \citep{pober2014next}, simulating noise using the auto-correlations as a measure of variance, and bootstrapping across the many possible pairs of baselines. 
As this last step is computationally expensive, scaling with the amount of $uv$ space analyzed, we limited our baseline selection to three vectors of length $\sim28\,m$, of differing orientation. 
These are the shortest baselines included in the mainline analysis. 
This restriction to a narrow $uv$ range is the largest divergence from the main power spectrum processing.

As the alternate pipeline was not developed in close proximity to the simulation, we were afforded an opportunity to trial blind testing. 
A blind test using realistic simulations provides an opportunity to test our judgement in identifying whether data points that are not noise-dominated arise from foregrounds, systematics or true 21\,cm signal.

A small subgroup, disconnected from the main Validation team, and blind to the preparation of the mock data, were set a challenge in which they were to distinguish between two datasets that were the same in every respect, except that one had 21\,cm signal and the other did not. 
These simulation products were blinded by changing filenames and removing metadata, and provided to the shadow-pipeline team after the ``Coherent Time Averaging'' step (cf. \autoref{fig:validation_pipe}). 

\begin{figure*}
    \centering
    \includegraphics[width=\linewidth,trim=2cm 0cm 3cm 0cm,clip=true]{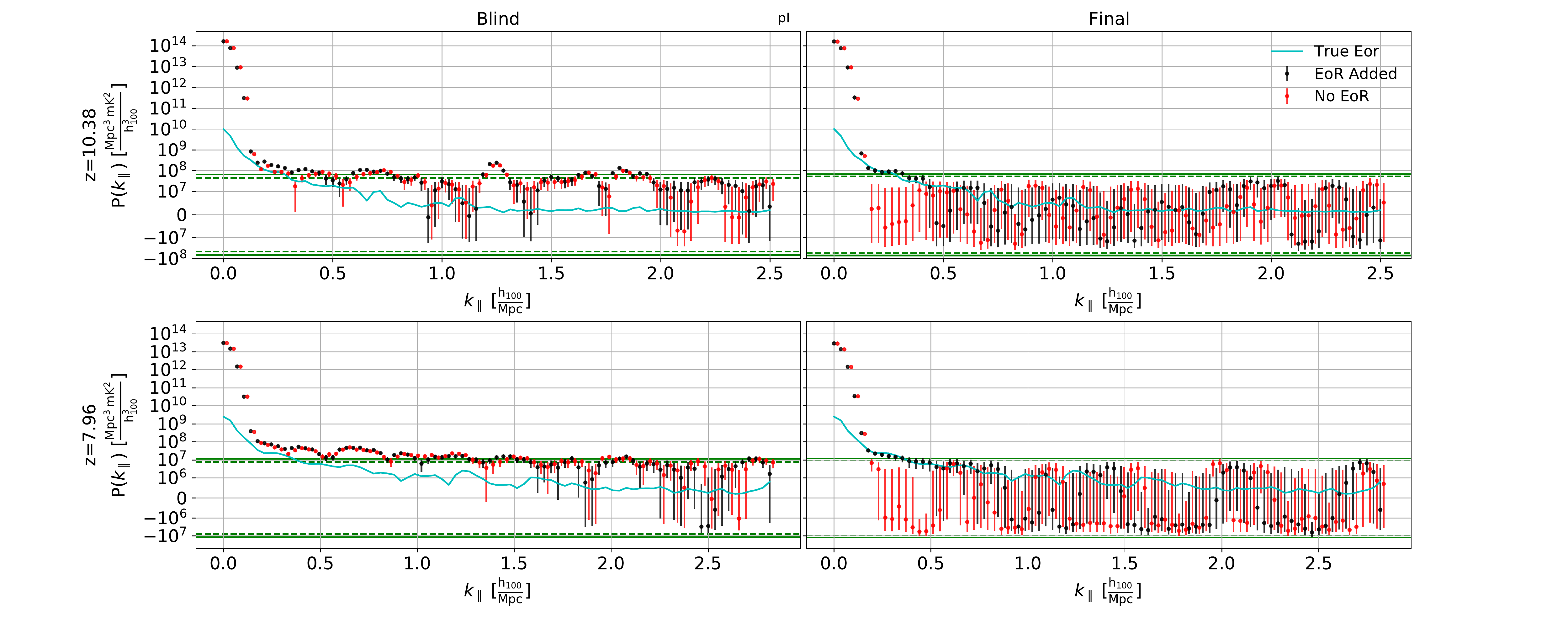}
    \caption{Two blinded simulations were processed using an independently developed shadow pipeline (\emph{simpleDS}, \citet{Kolopanis2019}). One contained a detectable 21cm signal. The initial result by a completely blinded team (left) was dominated by a strong systematic common to both simulations. After carefully limited discussion between the simulation team and the shadow group about weighting and data selection, but not unblinding the files, a strong distinction emerged (right) which the group interpreted as evidence for a (simulated) 21cm signal. This was confirmed in unblinding by adding the expected 21cm signal. Green lines show theoretically predicted noise power ($P_N$) at the 1- and 2-$\sigma$ level.}
    \label{fig:blind_test}
\end{figure*}

Fig. \ref{fig:blind_test} summarizes the results. 
In the first analysis, no cosmological signal could be clearly identified in the data (cf. left panel of \autoref{fig:blind_test}). 
Residuals were strong enough to make all data sets look roughly the same. 
Having finalized and reported this blind result to the rest of the Validation team, a meeting was held in which the topics discussed were intentionally limited to comparison of data selection between pipelines, and some clarification of the meaning of certain metadata. Importantly, the form and amplitude of the 21\,cm signal were kept hidden.
Each change discussed during these conversations was recorded and tested one at a time. 
The final resulting power spectrum estimate, obtained as a result of these limited discussions, is shown in the right column of \autoref{fig:blind_test}.
This figure shows clear improvement over the fully-blinded analysis shown in the left column, and indeed confirms that the alternate pipeline is able to accurately detect the input signal, and differentiate that detection from data without the signal.

The largest improvement between the left and right panels of \autoref{fig:blind_test} came from correctly interpreting sample-count metadata. 
The main analysis pipeline assigns flagged channels a sample count of zero, but then in-paints some of these channels (as described in \S\ref{sec:Methods:Flagging}). 
These in-painted data points are meant to be used when computing the delay spectra, but not to contribute towards the estimation of noise. 
The alternate pipeline was erroneously re-flagging these channels, due to a misunderstanding of the intention of the sample count.
With this misinterpretation, the flagged channels produced a large level of ``ringing'' when Fourier transformed, appearing as anomalous power in the estimated power spectrum.

Despite the aforementioned meeting to obtain clarification on the metadata, some differences between the pipelines were intentionally maintained. The single biggest difference between the pipelines was the selection of the LST range, which was smaller for the shadow pipeline (about an hour shorter).
Furthermore, the shadow analysis averaged both fields together. Clearly, these differences do not significantly affect the conclusions of the test. Indeed, they further strengthen the case that the analysis is not highly sensitive to the precise choice of LST range.

This test increases confidence in the power spectrum portion of the analysis, re-enforces the conclusion that loss within the calibration pipeline is minimal, and provides a guide for how blind comparison between simulation and data can be employed to assess the relative likelihood that an observed power level is due to a true background.

\subsection{Accuracy of Error Bars}

The level of the injected EoR power spectrum is such that there are four regimes at the final noise level (integrating all times and baselines): foreground dominated for $k < 0.2$, EoR dominated for $0.2 < k < 0.4$, systematics dominated (before subtraction) for $0.4 < k < 0.55$ and noise dominated for $k > 0.6$.  Consequently, we can assess the consistency of the recovered data points with the error bars in a manner similar to that of \HOneClimit\ (Equation 26 and Table 5).  The null hypothesis here is that the data points are consistent with zero, given the error bars reported.  
Performing this test, we find a significant detection ($p < 0.001$) in all bands and fields when including all $k > 0.2$, consistent with detections of the injected EoR.  For $k > 1$, all bands and fields had a $p$-value consistent with the null hypothesis, except for Band 2, Field 2, which has two 2-$\sigma$ outliers at the highest $k$.  The imaginary component of the power spectrum is consistent with the null hypothesis for all $k > 0.2$ and all bands and fields.

\section{Discussion and Conclusions} 
\label{sec:Conclusions}

\subsection{Basic conclusions}

In general, we have found that the HERA H1C software pipeline successfully reproduces known analytic input power spectra, under the assumptions it adopts; we did not find major issues with any of the pipeline steps we investigated here.

We performed power spectrum estimation with a full end-to-end mock dataset including a wide range of realistic instrumental effects and foregrounds (c.f. \S\ref{sec:EndToEnd}).
In this test, mock visibility data were self-consistently generated from a known analytic power spectrum (c.f. \S\ref{sec:mock_eor_step0} and \S\ref{sec:meth:simulators}), obscured with realistic Galactic and extra-galactic foreground models (c.f. \S\ref{sec:Methods:ForegroundModels}), and contaminated with almost all known instrumental effects relevant to the HERA instrument (c.f. \S\ref{sec:meth:noise}).
A summary of the included components can be found in \autoref{fig:validation_matrix} and the top panel of \autoref{tab:WhatsInAndOut}.
This mock data, simulated to be broadly consistent with H1C observing parameters, was passed through the full H1C analysis and power spectrum estimation pipeline, with all analysis parameters consistent with those used for processing actual data (c.f. \HOneClimit).
As our primary result, we demonstrated that the pipeline produces power spectrum estimates that are consistent with the known analytic input to within thermal noise levels (at the $2\sigma$ level) for $k > 0.2 h{\rm Mpc}^{-1}$ for both bands and fields considered (c.f. \autoref{fig:test4_lstcuts}).

To test the pipeline in various regimes in which different components dominate, the analytic input spectrum was intentionally amplified to enable a strong ``detection'' at $k\sim 0.2\,h{\rm Mpc}^{-1}$ -- at the level of $\sim 25\sigma$ -- with foregrounds dominating on larger scales, thermal noise dominating at smaller scales, and systematics dominating (before subtraction) in between.
The pipeline successfully detected this amplified input signal, after suppressing foregrounds with a dynamic range (foreground to noise ratio) of $\gtrsim 10^7$. 
Additionally, the noise-dominated power spectrum at high $k$ was found to be consistent with the predicted noise power.
With the possible exception of a single $k$-bin in ``Band 1, Field 2'', systematics were mitigated to below the noise level of the simulation.

This does not guarantee that there are no inaccuracies remaining, but we can be confident that any are unlikely to have major effects on the \HOneClimit\ results.
Recall that the goal of this effort was to validate the {\it software and algorithms}, not the {\it data}.   
Thus, there may yet be subtle effects present in the real data which we did not adequately represent in the simulation, or analysis choices which do not perform correctly when the assumptions of the pipeline are violated.  
However, substantial issues like those found in \cite{Cheng2018}, which were the result of the algorithm, independent of the data, seem unlikely.

A number of small problems and unanticipated effects were discovered (e.g., those given in \cref{tab:signal_loss}) and these have led either to improvements in the existing pipeline which eliminate them, or inclusion in a list of effects to continue investigating.

\subsection{Scope of Future Work}

\begin{table}[h]
    \centering
    \begin{tabular}{l}
        \hline
        \textbf{\textsc{Simulated Effects}} \\
        \hline
        Known bright point sources \\
        Point-source foregrounds from GLEAM catalog \\ 
        Diffuse foregrounds on scales $>3\arcdeg$ \\ 
        Sky-derived thermal noise \\
        Simple receiver noise \\
        Realistic EM-simulated beam \\
        Direction-independent gains \\ 
        ~~~(per feed, time invariant) \\
        Cross-coupling model \\
        Per-antenna cable reflections \\
        Realistic flagging patterns \\
        \hline
        \textbf{\textsc{Neglected Effects (Major)}} \\
        \hline
        Mis-identification of dysfunctional antennas \\
        Mis-identification of RFI \\
        Generation of \texttt{abscal} model from images \\
        Antenna non-redundancy (in primary beams) \\
        Fully realistic antenna-cross coupling \\
        Antenna position errors \\
        \hline
        \textbf{\textsc{Neglected Effects (Minor)}} \\
        \hline
        Digital (correlator) artefacts \\
        Confused point sources \\
        Fully polarized direction-independent gains \\
        Time variation in direction-independent gains \\
        ~~~(due to, e.g., temperature variations) \\
        Polarized and/or transient sources \\
        Full suite of possible shapes for $P_{\rm EoR}(k)$ \\
        Ionosphere \\
        \hline 
    \end{tabular}
    \caption{Effects included and not included in this analysis.
    }
    \label{tab:WhatsInAndOut}
\end{table}

As the HERA pipeline improves and changes, the validation effort will need to continue to include simulations which effectively test the new software and challenge the assumptions made.  There are some obvious axes along which the validation effort will need to be extended or modified for future work; we briefly list some of them here.

In this work, we have compared portions of the analysis between two pipelines: the HERA standard one and \simpleDS.  This comparison was done nearly ``blind'', i.e., the group analyzing using \simpleDS\ did not know anything about the datasets that were prepared for it, and analyzed the data as if were from the real instrument.  Both aspects of this cross-check should be kept in future validation efforts, namely the existence of a parallel pipeline, and the independent, blind analysis of the simulated datasets.  This turns up both differences due to the different algorithms, but is also revealing of different implicit assumptions in the analyses.

A more complete simulation of RFI and the effectiveness of the flagging is clearly essential as the limits get deeper to ensure that there is no significant effect due to unflagged RFI.  
In part, this requires devising and implementing a suite of null tests on the real data, since the complexity of actual RFI will probably always exceed our ability to simulate it.  
Nevertheless, it should be possible to gain considerably more insight via a detailed simulation step into how well our current algorithms are doing and whether there are likely gaps in their effectiveness 
\citep{hera_memo_82}.

The simulations here have several aspects that are specific to the instrument configuration.  An entirely new feed system is currently being commissioned \citep{deleraacedo20antenna, fagnoni20}, which will necessitate a new set of investigations of systematic effects, new simulations of them, and tests of the methods proposed to correct or mitigate them.

The continued improvement in our understanding of foregrounds, including both better point source catalogs and better models of diffuse emission, particularly below 100 MHz, will be folded into future validation simulations.  
The foreground models should also be broadened to include Faraday-rotated polarized emission, to simulate the effects of polarized leakage which could be comparable to the level of EoR, especially for the delay spectrum approach, which does not segregate the polarized signal via an image-based analysis \citep{Nunhokee2017, asad16, asad17}.  
The effects of the ionosphere may also be important, particularly its interaction with the polarized foregrounds \citep{Martinot2018}.  

Another complication not directly addressed by our analysis is the impact of an incomplete or incorrect sky and/or beam model on the HERA absolute calibration step. In addition to the flux-scale issues (quantified in \citet{Kern2020_abscal} and accounted for in \HOneClimit), this effect can introduce spurious spectral structure into the calibration solutions \citep{Barry2016,Byrne2019} which can be mitigated by including only short baselines \citep{EwallWice2017, Orosz2019} in the calibration. Because the reference visibilities used in absolute calibration differ little from the true visibilities (they are filtered at high delay as explained in \autoref{subsec:HERAPipeline}), the impact of sky-model error cannot be quantified here. However, because calibration solutions are smoothed spectrally at delays larger than 100\,ns, we avoid spectral structure from modeling error by simply not trying to calibrate any true spectral structure in the instrument response beyond 100\,ns unless it can be modeled as the sum of reflections and inferred from the auto-correlations. While this smoothing was primarily motivated by the desire to mitigate the effect of cross-coupling on the gains \citep{Kern2020_abscal}, it makes the spectral impact of modeling error largely irrelevant for this analysis. If additional degrees of freedom are admitted in the calibration solutions in the future (e.g.\ by increasing the delay threshold for smoothing), this question needs to be revisited.

Such considerations also point to the complex open question of how to simulate the effects of violations of the assumptions of the analysis, particularly as regards incomplete knowledge of the primary beams of the antennas and failure of redundancy between nominally redundant baselines for various reasons.  
The ability to simulate a different primary beam for each antenna is included in our simulation packages (\texttt{RIMEz} and  \texttt{pyuvsim}), but how to represent realistic variations is currently a topic of active research \citep[eg.][]{Choudhuri2021}.  
Another example of this is simulating time-variable gains which capture the actual instrument behavior.

Another consideration not addressed in this validation suite, but firmly in place for future tests, is the applicability of the pipeline to markedly different (but physically reasonable) shapes for $P_{\rm eor}(k)$.
For instance, one might imagine that a ``sharp'' feature in $P_{\rm eor}(k)$ might cause difficulties for power spectrum estimation.  The interaction of the window functions with the power spectrum also needs to be considered more carefully.

Finally, the end-to-end approach here only considered a simulation of a single dataset of approximately the same size as the actual one.
As the HERA data grows, this will be an increasingly difficult task, to say nothing of the need for exploring errors via multiple realizations of noise, systematic effects, and cosmological signal.  
In particular, our criteria for what constitutes a successful end-to-end test will need to be more rigorously tied to keeping systematic errors from the analysis to less than the random errors due to instrument noise (and its coupling to signal), combined with the expected cosmic variance \citep[e.g.,][]{Lanman2019}.
This will also have an effect on how we assess the errors on tests of portions of the pipeline (the ``steps'' in \cref{fig:validation_matrix}).
We will need to investigate further which aspects of the pipeline truly require a simulated dataset comparable to the full one, and which require multiple realizations to understand the statistical effects.  
This is particularly important with respect to systematic effects whose exact parameterization is difficult to quantify (e.g., primary beam non-redundancy).  


\section*{Acknowledgements}
This material is based upon work supported by the National Science Foundation under Grant Nos. 1636646 and 1836019 and institutional support from the HERA collaboration partners.
This research is funded in part by the Gordon and Betty Moore Foundation.
HERA is hosted by the South African Radio Astronomy Observatory, which is a facility of the National Research Foundation, an agency of the Department of Science and Innovation.
Parts of this research were supported by the Australian Research Council Centre of Excellence for All Sky Astrophysics in 3 Dimensions (ASTRO 3D), through project number CE170100013.
J.~Aguirre acknowledges funding from NSF CAREER grant AST-1455151.
G.~Bernardi acknowledges funding from the INAF PRIN-SKA 2017 project 1.05.01.88.04 (FORECaST), support from the Ministero degli Affari Esteri della Cooperazione Internazionale - Direzione Generale per la Promozione del Sistema Paese Progetto di Grande Rilevanza ZA18GR02 and the National Research Foundation of South Africa (Grant Number 113121) as part of the ISARP RADIOSKY2020 Joint Research Scheme, from the Royal Society and the Newton Fund under grant NA150184 and from the National Research Foundation of South Africa (grant No. 103424).
P.~Bull acknowledges funding for part of this research from the European Research Council (ERC) under the European Union's Horizon 2020 research and innovation programme (Grant agreement No. 948764), and from STFC Grant ST/T000341/1.
J.~S.~Dillon gratefully acknowledges the support of the NSF AAPF award \#1701536.
N.~Kern acknowledges support from the MIT Pappalardo fellowship.
A.~Liu acknowledges support from the New Frontiers in Research Fund Exploration grant program, the Canadian Institute for Advanced Research (CIFAR) Azrieli Global Scholars program, a Natural Sciences and Engineering Research Council of Canada (NSERC) Discovery Grant and a Discovery Launch Supplement, the Sloan Research Fellowship, and the William Dawson Scholarship at McGill.
M.~G.~Santos acknowledges support from the South African Square Kilometre Array Project and National Research Foundation (Grant No. 84156).

\section*{Contributions of the Authors}
\label{sec:AuthorContributions}

JEA leads the HERA Validation group and was responsible for the overall direction and organization of the validation effort presented here.  SGM created and curates the {\tt hera-validation} repository, directed validation of the visibility simulators (Step -1.1) and ran the foreground tests (Steps 1.1 and 1.2).  RP created the end-to-end simulations and ran the Step 0.1 and 2.1 tests.  JB tested the in-painting and systematic subtraction in Step 3.1.  JSD ran the calibration simulations/pipelines and their tests (Steps 2.0 and 4).  MK and DCJ analyzed the blind test. DCJ wrote the results of the blind test and outside simulator validation.
NSK ran the power spectrum pipeline on the end-to-end simulations and performed the signal loss tests (\autoref{sec:EndToEnd}).
LW ran the foreground tests (Step 1.0) and the \rimez\ validation (Step -1.1).
The HERA Builder's list is included alphabetically as authors because of the dependence of this work on the combined efforts of collaborators on the various software repositories as well as the necessity of using HERA data to build the models used in this paper.

\bibliography{bibliography}

\appendix

\section{The HERA Validation Subsystem}
\label{sec:ValidationSubsystem}

The HERA collaboration has placed a high emphasis on detailed validation by establishing a dedicated Validation team, formalized as an essential HERA subsystem.
HERA `subsystems' are the major components of the HERA experiment, and have a dedicated team associated with each.
In addition to the ``Validation'' subsystem, others include ``Power Spectrum Estimation'', ``Analysis'', ``Quality Metrics'' and ``Inclusion/Diversity''.

The scope of this effort is clearly wide-ranging: ultimately it is to verify that the  reported power spectra from the HERA collaboration are free from defect, whether from code bugs, poor algorithmic choices, or inappropriate physical assumptions.
At the same time, the goal is not to merely internally validate, but also to ensure that the pipeline is reproducible and understandable by the wider community, in order to build confidence in reported upper limits or detections.

\subsection{Code Standards}
\label{sec:CodeStandards}

HERA has adopted a set of high-standard open source software practices which encourage transparency, reproducibility, interoperability and peer-verification.
All systems-level HERA code is hosted open-source on a single GitHub organization\footnote{\url{https://github.com/hera-team} -- note that not all repositories found here are considered `systems-level'.}.
A set of well-defined software standards are applicable across the organization, encouraging a certain degree of homogeneity between project-level packages.  
Amongst these standards are 
\begin{itemize}
    \item Documentation: Python code is self-documented (i.e. includes `docstrings'\footnote{Docstrings are a Python construct for documenting code objects in-place in the code, and can be used to automatically create up-to-date online documentation.} for all public modules, functions, classes and methods), using a uniform docstring format (typically \textsc{numpydoc}). Extra tutorials and examples are also encouraged.
    \item Testing: all systems-level HERA packages are thoroughly unit-tested\footnote{Unit-tests are functions that assert specific conditions on the behaviour of the \textit{basic units} of the software (eg. functions or class methods), and can be collected and run together in an automated fashion. This is in contrast to \textit{integration tests} which assert conditional behaviour of combinations of the basic units.}, and kept at $>95\%$ code-coverage\footnote{Code coverage represents the percentage of standard lines of code in the package that are run during execution of the test suite.}. Testing is performed continuously via an online Continuous Integration provider (e.g. Travis or Github Actions).
    \item Formatting: all code is \textsc{pep8}-compliant\footnote{\url{https://www.python.org/dev/peps/pep-0008/}} (often enforced by the use of external tools such as \textsc{black}\footnote{\url{https://black.readthedocs.io}} \textsc{pre-commit}\footnote{\url{https://pre-commit.com/}}), making each package more homogeneous (important when there are many contributors to the repository) and easy to read. This is important for transparency both within and without the collaboration.
    \item Review: each package uses the GitHub flow\footnote{\url{https://guides.github.com/introduction/flow/}} as a software delivery workflow. In brief, in this workflow the \textit{master}\footnote{In the near future, the \texttt{master} branch will be renamed to \texttt{main}, as is now widely endorsed.} branch is considered protected, and is disabled for direct code changes on GitHub. This requires new code additions (and bug fixes) to be developed in a branch that is \textit{not master}, and a formal \textit{pull request} (PR) to be created and accepted before merging back into the protected \textit{master} branch. All repositories have an option enabled in which PRs must be first reviewed and accepted by a person other than the author before they can be merged. PRs must also satisfy a host of other status checks, such as passing Continuous Integration tests and satisfying coverage checks. 
    Such reviews lessen the probability that subtle bugs enter the code (especially those that are only apparent when one has familiarity with a \textit{different} part of the code), but also serve to increase the overall familiarity with the code base, as it evolves, of the wider collaboration.
\end{itemize}

\subsection{The Validation Code Repository}
\label{sec:ValidationRepo}

Following the lead of the wider collaboration, the HERA Validation team has established a public repository in which all pipeline validation tests are performed and archived.\footnote{\url{https://github.com/hera-team/hera-validation}}
We have defined a comprehensive set of tests of the pipeline, moving from simplistic analyses through to a full end-to-end simulation and analysis (cf. \S\ref{sec:meth:steps}). 
Each of these tests is performed and documented in a Jupyter notebook,\footnote{\url{https://jupyter.org}} developed and archived on our GitHub repository. 
Jupyter notebooks allow combining arbitrary documentation and code execution in order to generate a full analysis. 
We utilise this ability, adopting a certain template for each test that includes listing the full provenance of all data used, the exact package versions of all dependent software (down to the git hash), a summary description, a set of criteria to meet for the test to pass, and a list of suggestions for follow-up tests.
These sections promote reproducibility and clarity.

Each test is recorded via a three-digit identifier: \verb|major.minor.test|, in which the \verb|major| digit identifies a broad class of physical effects being tested, the \verb|minor| digit identifies variations on that class of physical effects, and the \verb|test| digit represents an iteration in the testing procedure (eg. a test may fail and require re-running with a bugfix, or with a slight alteration in the assumptions).
Each notebook contains a single test. 
Although all tests are version-controlled, we do \textit{not} overwrite test notebooks when an updated test is performed. The failed or outdated tests are kept at the top-level of the repository to make it easy to determine the history of the test.\footnote{Jupyter notebooks are also not particularly well-suited for granular version control.}

In keeping with the standards of the rest of the collaboration, validation tests are required to be reviewed and accepted by the rest of the group before being merged into the master branch.

Extra features of GitHub have also been used to aide in the organization of the Validation effort. In particular, newly proposed tests are created as GitHub issues, where they are discussed before accepting them into the test-suite canon. A set of custom tags has been specified, explicitly defining each simulation and analysis component the test would validate (cf. Figure \ref{fig:validation_matrix}).

This system has served well in this particular validation effort, and will continue to be used to develop further validation tests for upcoming data releases. 

\section{Window Functions and Aliasing}
\label{sec:WindowFunction}

This appendix seeks to explain the discrepancy between the analytic input and estimated power spectrum, and present the definition of the ``aliased" power spectrum in \autoref{fig:PowerSpectrumValidation} and why it is much closer to the estimated power spectrum. 

In general the power spectrum estimates $\widehat{P}(k)$ produced with \herapspec\ in this paper can be described by
\begin{align}
\widehat{P}(k) = \sum_{\alpha} \sum_{\beta} E(k, \alpha, \beta) V(\alpha) V^*(\beta)
\end{align}
where $V$ is the visibility function and $\alpha$, $\beta$ are indices over the set of points $\alpha = (t, \nu, \va{b})$ at which the visibility function is measured. Let the visibility $V$ be sourced by only the cosmological signal, as in the simulation that produces \autoref{fig:PowerSpectrumValidation}. 
The covariance matrix of the data is then a linear functional of the power spectrum, which may be written 
\begin{align}
\expval{ V(\alpha) V^*(\beta) } & = \int_{0}^{\infty} \pdv{\expval{ V(\alpha) V^*(\beta) } }{P(k)} P(k) \dd{k}.
\end{align}
The expectation value of the power spectrum estimate is thus
\begin{align}
\expval{\widehat{P}(k)} & = \sum_{\alpha \beta} E(k, \alpha, \beta) \expval{V(\alpha) V^*(\beta)} \\
& = \int_{0}^{\infty} \sum_{\alpha \beta} E(k, \alpha, \beta) \pdv{\expval{ V(\alpha) V^*(\beta) } }{P(k')} P(k') \dd{k'}.
\end{align}
We define the ``window function''
\begin{align}
W(k, k') \equiv \sum_{\alpha \beta} E(k, \alpha, \beta) \pdv{\expval{ V(\alpha) V^*(\beta) } }{P(k')}, 
\label{eq:GeneralWindow}
\end{align}
and hence an unbiased estimator for the power spectrum -- i.e. an estimator such that
\begin{align}
\expval{\widehat{P}(k)} & = P(k)
\end{align}
-- would be one such that the window function is 
\begin{align}
\sum_{\alpha \beta} E(k, \alpha, \beta) \pdv{\expval{ V(\alpha) V^*(\beta)}}{P(k')} & = \delta(k - k').
\end{align}
With a countable number of samples $\alpha$ and finite bandwidths of our measurements it is not possible to achieve such a window function exactly -- any real measurement will be ``corrupted" by a window function $W(k, k')$ with a finite width. 
The best that can be accomplished is that the window function is localized, so that the estimate of $\widehat{P}(k)$ has contributions from only $k'$ nearby to $k$. The delay spectrum estimator applied in  \autoref{fig:PowerSpectrumValidation} has the form
\begin{align}
E(k,t, \nu_n, \va{b}_i, t', \nu_m, \va{b}_j) & =  \mathcal{N} w_n w_m \exp(i \mathcal{A} k (n - m)) \delta_{t, t'} \delta_{ij}
\end{align}
(i.e. $\mathcal{N}$ is the delay spectrum normalization scalar and $\mathcal{A}$ is the frequency band dependent delay conversion factor)  which produces a localized window function when $w_n$ is a frequency taper like the Blackmann-Harris used in our power spectrum estimates. 

It is this effect that causes the evident discrepancy between the input and estimated power spectrum in \autoref{fig:PowerSpectrumValidation}. In the low-$k$ regime ($k \sim 0.1$) the window function has an approximately constant width (in linear units of $k$), except for the lowest several $k$ points. Each power spectrum estimate is an integral over the true power spectrum within that constant width, but as the estimated $k$ point decreases toward low-$k$, the intrinsic power-law power spectrum increasingly varies over the width of the window function. This causes the estimate to be increasingly biased high with respect to the central value of the analytic input. However at the lowest two or three $k$ points the window functions become much less well behaved i.e more oscillatory and less localized, which produces the dip in the lowest $k$ point in \autoref{fig:PowerSpectrumValidation}.

At high $k$ the effect of the window function is approximated by a classical aliasing calculation -- aliasing in a DFT-based power spectrum estimate is described by a window function, and in this case the aliasing window function is a decent approximation of the true window function in our test. To see this, consider the window function induced by the classical aliasing in a DFT and the resulting power spectrum. If a Gaussian random function $f(r)$ has a power spectrum $P(k)$ i.e.
\begin{align}
f(r) & = \int_{-\infty}^\infty \frac{\dd{k}}{2 \pi} e^{2 \pi i k r} \tilde{f}(k) \\
\expval{ \tilde{f}(k) \tilde{f}^(k') } & = 2 \pi P(k) \delta(k - k')
\end{align}
and $f(r)$ is sampled at a rate $2 k_s$ with samples $f_n = f(r_n)$ and the power spectrum is estimated from the DFT estimate
\begin{align}
\tilde{f}(k_m) \propto \sum_{n=1}^{N} e^{2 \pi i \frac{n}{N} m} f_n
\end{align}
then the measured power spectrum $\hat{P}(k_m) = \abs{\tilde{f}(k_m)}^2$ has an expectation value $\expval{\hat{P}} \approx P_{\rm aliased}$ (for $N \sim > 100$) where, according to the well-known equation,
\begin{align}
\label{eq:AliasedPk}
P_{\rm aliased}(k) & = P(k) + \sum_{n=1}^\infty P(2 n k_s - k) + P(2 n k_s + k).
\end{align}
This can be expressed as the effect of a window function defined as
\begin{align}
W(k, k') & = \delta(k - k')  + \sum_{n=1}^{\infty} \delta(2 n k_s - k - k') + \delta(2 n k_s + k - k')
\label{eq:Aliasing}
\end{align}
and then for $k \in [0, k_s)$
\begin{align}
P_{aliased}(k) = \int_0^\infty W(k, k') P(k') \dd{k'}.
\end{align}

The function $P_{\rm aliased}$ is the ``aliased" power spectrum in \autoref{fig:PowerSpectrumValidation}.

\end{document}

%% file: author-list.tex
\correspondingauthor{J.~Aguirre}
\email{jaguirre@sas.upenn.edu}

\author{James E. Aguirre}
\affiliation{Department of Physics and Astronomy, University of Pennsylvania, Philadelphia, PA}

\author{Steven G. Murray}
\affiliation{School of Earth and Space Exploration, Arizona State University, Tempe, AZ}

\author{Robert  Pascua}
\affiliation{Department of Astronomy, University of California, Berkeley, CA}

\author{Zachary E. Martinot}
\affiliation{Department of Physics and Astronomy, University of Pennsylvania, Philadelphia, PA}

\author{Jacob  Burba}
\affiliation{Department of Physics, Brown University, Providence, RI}

\author{Joshua~S.~Dillon}
\altaffiliation{NSF Astronomy and Astrophysics Postdoctoral Fellow}
\affiliation{Department of Astronomy, University of California, Berkeley, CA}

\author{Daniel C. Jacobs}
\affiliation{School of Earth and Space Exploration, Arizona State University, Tempe, AZ}

\author{Nicholas S. Kern}
\affiliation{Department of Physics, Massachusetts Institute of Technology, Cambridge, MA}

\author{Piyanat  Kittiwisit}
\affiliation{Department of Physics and Astronomy,  University of Western Cape, Cape Town, 7535, South Africa}

\author{Matthew  Kolopanis}
\affiliation{School of Earth and Space Exploration, Arizona State University, Tempe, AZ}

\author{Adam Lanman}
\affiliation{Department of Physics and McGill Space Institute, McGill University, 3600 University Street, Montreal, QC H3A 2T8, Canada}

\author{Adrian  Liu}
\affiliation{Department of Physics and McGill Space Institute, McGill University, 3600 University Street, Montreal, QC H3A 2T8, Canada}

\author{Lily Whitler}
\affiliation{School of Earth and Space Exploration, Arizona State University, Tempe, AZ}

\author{Zara  Abdurashidova}
\affiliation{Department of Astronomy, University of California, Berkeley, CA}

\author{Paul  Alexander}
\affiliation{Cavendish Astrophysics, University of Cambridge, Cambridge, UK}

\author{Zaki S. Ali}
\affiliation{Department of Astronomy, University of California, Berkeley, CA}

\author{Yanga  Balfour}
\affiliation{South African Radio Observatory (SARAO), 2 Fir Street, Observatory, Cape Town, 7925, South Africa}

\author{Adam P. Beardsley}
\affiliation{School of Earth and Space Exploration, Arizona State University, Tempe, AZ}

\author{Gianni  Bernardi}
\affiliation{Department of Physics and Electronics, Rhodes University, PO Box 94, Grahamstown, 6140, South Africa}
\affiliation{INAF-Istituto di Radioastronomia, via Gobetti 101, 40129 Bologna, Italy}
\affiliation{South African Radio Observatory (SARAO), 2 Fir Street, Observatory, Cape Town, 7925, South Africa}

\author{Tashalee S. Billings}
\affiliation{Department of Physics and Astronomy, University of Pennsylvania, Philadelphia, PA}

\author{Judd D. Bowman}
\affiliation{School of Earth and Space Exploration, Arizona State University, Tempe, AZ}

\author{Richard F. Bradley}
\affiliation{National Radio Astronomy Observatory, Charlottesville, VA}

\author{Philip Bull}
\affiliation{School of Physics \& Astronomy, Queen Mary University of London, London, UK}
\affiliation{Department of Physics and Astronomy,  University of Western Cape, Cape Town, 7535, South Africa}

\author{Steve  Carey}
\affiliation{Cavendish Astrophysics, University of Cambridge, Cambridge, UK}

\author{Chris L. Carilli}
\affiliation{National Radio Astronomy Observatory, Socorro, NM}

\author{Carina  Cheng}
\affiliation{Department of Astronomy, University of California, Berkeley, CA}

\author{David R. DeBoer}
\affiliation{Department of Astronomy, University of California, Berkeley, CA}

\author{Matt  Dexter}
\affiliation{Department of Astronomy, University of California, Berkeley, CA}

\author{Eloy  de~Lera~Acedo}
\affiliation{Cavendish Astrophysics, University of Cambridge, Cambridge, UK}

\author{John  Ely}
\affiliation{Cavendish Astrophysics, University of Cambridge, Cambridge, UK}

\author{Aaron  Ewall-Wice}
\affiliation{Department of Astronomy and Physics, University of California, Berkeley, CA}

\author{Nicolas  Fagnoni}
\affiliation{Cavendish Astrophysics, University of Cambridge, Cambridge, UK}

\author{Randall  Fritz}
\affiliation{South African Radio Observatory (SARAO), 2 Fir Street, Observatory, Cape Town, 7925, South Africa}

\author{Steven R. Furlanetto}
\affiliation{Department of Physics and Astronomy, University of California, Los Angeles, CA}

\author{Kingsley  Gale-Sides}
\affiliation{Cavendish Astrophysics, University of Cambridge, Cambridge, UK}

\author{Brian  Glendenning}
\affiliation{National Radio Astronomy Observatory, Socorro, NM}

\author{Deepthi  Gorthi}
\affiliation{Department of Astronomy, University of California, Berkeley, CA}

\author{Bradley  Greig}
\affiliation{School of Physics, University of Melbourne, Parkville, VIC 3010, Australia}

\author{Jasper  Grobbelaar}
\affiliation{South African Radio Observatory (SARAO), 2 Fir Street, Observatory, Cape Town, 7925, South Africa}

\author{Ziyaad  Halday}
\affiliation{South African Radio Observatory (SARAO), 2 Fir Street, Observatory, Cape Town, 7925, South Africa}

\author{Bryna J. Hazelton}
\affiliation{Department of Physics, University of Washington, Seattle, WA}
\affiliation{eScience Institute, University of Washington, Seattle, WA}

\author{Jacqueline N. Hewitt}
\affiliation{Department of Physics, Massachusetts Institute of Technology, Cambridge, MA}

\author{Jack  Hickish}
\affiliation{Department of Astronomy, University of California, Berkeley, CA}

\author{Austin  Julius}
\affiliation{South African Radio Observatory (SARAO), 2 Fir Street, Observatory, Cape Town, 7925, South Africa}

\author{Joshua  Kerrigan}
\affiliation{Department of Physics, Brown University, Providence, RI}

\author{Saul A. Kohn}
\affiliation{Department of Physics and Astronomy, University of Pennsylvania, Philadelphia, PA}

\author{Paul  La~Plante}
\affiliation{Department of Physics and Astronomy, University of Pennsylvania, Philadelphia, PA}

\author{Telalo  Lekalake}
\affiliation{South African Radio Observatory (SARAO), 2 Fir Street, Observatory, Cape Town, 7925, South Africa}

\author{David  Lewis}
\affiliation{School of Earth and Space Exploration, Arizona State University, Tempe, AZ}

\author{David  MacMahon}
\affiliation{Department of Astronomy, University of California, Berkeley, CA}

\author{Lourence  Malan}
\affiliation{South African Radio Observatory (SARAO), 2 Fir Street, Observatory, Cape Town, 7925, South Africa}

\author{Cresshim  Malgas}
\affiliation{South African Radio Observatory (SARAO), 2 Fir Street, Observatory, Cape Town, 7925, South Africa}

\author{Matthys  Maree}
\affiliation{South African Radio Observatory (SARAO), 2 Fir Street, Observatory, Cape Town, 7925, South Africa}

\author{Eunice  Matsetela}
\affiliation{South African Radio Observatory (SARAO), 2 Fir Street, Observatory, Cape Town, 7925, South Africa}

\author{Andrei  Mesinger}
\affiliation{Scuola Normale Superiore, 56126 Pisa, PI, Italy}

\author{Mathakane  Molewa}
\affiliation{South African Radio Observatory (SARAO), 2 Fir Street, Observatory, Cape Town, 7925, South Africa}

\author{Miguel F. Morales}
\affiliation{Department of Physics, University of Washington, Seattle, WA}

\author{Tshegofalang  Mosiane}
\affiliation{South African Radio Observatory (SARAO), 2 Fir Street, Observatory, Cape Town, 7925, South Africa}

\author{Abraham R. Neben}
\affiliation{Department of Physics, Massachusetts Institute of Technology, Cambridge, MA}

\author{Bojan  Nikolic}
\affiliation{Cavendish Astrophysics, University of Cambridge, Cambridge, UK}

\author{Aaron R. Parsons}
\affiliation{Department of Astronomy, University of California, Berkeley, CA}

\author{Nipanjana  Patra}
\affiliation{Department of Astronomy, University of California, Berkeley, CA}

\author{Samantha  Pieterse}
\affiliation{South African Radio Observatory (SARAO), 2 Fir Street, Observatory, Cape Town, 7925, South Africa}

\author{Jonathan C. Pober}
\affiliation{Department of Physics, Brown University, Providence, RI}

\author{Nima  Razavi-Ghods}
\affiliation{Cavendish Astrophysics, University of Cambridge, Cambridge, UK}

\author{Jon  Ringuette}
\affiliation{Department of Physics, University of Washington, Seattle, WA}

\author{James  Robnett}
\affiliation{National Radio Astronomy Observatory, Socorro, NM}

\author{Kathryn  Rosie}
\affiliation{South African Radio Observatory (SARAO), 2 Fir Street, Observatory, Cape Town, 7925, South Africa}

\author{Mario  G. Santos}
\affiliation{Department of Physics and Astronomy,  University of Western Cape, Cape Town, 7535, South Africa}
\affiliation{South African Radio Observatory (SARAO), 2 Fir Street, Observatory, Cape Town, 7925, South Africa}

\author{Peter  Sims}
\affiliation{Department of Physics and McGill Space Institute, McGill University, 3600 University Street, Montreal, QC H3A 2T8, Canada}
\affiliation{School of Earth and Space Exploration, Arizona State University, Tempe, AZ}

\author{Saurabh  Singh}
\affiliation{Department of Physics and McGill Space Institute, McGill University, 3600 University Street, Montreal, QC H3A 2T8, Canada}

\author{Craig  Smith}
\affiliation{South African Radio Observatory (SARAO), 2 Fir Street, Observatory, Cape Town, 7925, South Africa}

\author{Angelo  Syce}
\affiliation{South African Radio Observatory (SARAO), 2 Fir Street, Observatory, Cape Town, 7925, South Africa}

\author{Nithyanandan  Thyagarajan}
\affiliation{National Radio Astronomy Observatory, Socorro, NM}
\affiliation{Nithyanandan Thyagarajan is a Jansky fellow of the National Radio Astronomy Observatory.}
\affiliation{School of Earth and Space Exploration, Arizona State University, Tempe, AZ}

\author{Peter K.~G. Williams}
\affiliation{Center for Astrophysics | Harvard \& Smithsonian, Cambridge, MA}
\affiliation{American Astronomical Society, Washington, DC}

\author{Haoxuan  Zheng}
\affiliation{Department of Physics, Massachusetts Institute of Technology, Cambridge, MA}